\documentclass{svjour2}                    % onecolumn
\smartqed  % flush right qed marks, e.g. at end of proof
\usepackage{graphicx}

\usepackage{amsmath}
\usepackage{amssymb}
\usepackage{latexsym}
\usepackage{fancybox}
\usepackage{multicol}
\usepackage{pstricks}
\usepackage{pst-node}
\usepackage{pst-plot}
\usepackage{pst-tree}
\usepackage{subfigure}
\usepackage{natbib}
\usepackage{mathptmx}

\newpsobject{showgrid}{psgrid}{subgriddiv=1,griddots=10,gridlabels=6pt}

\def\ignore#1{}

\def\calS0{{\cal S}_0}

\def\cald{{\cal D}}
\def\calq{{\cal Q}}

\def\wnnw#1{\omega_{#1}}

\def\tc#1{TC({#1})}
\def\union#1#2{{#1}\cup{#2}}
\def\prior#1#2{{#1}\rhd{#2}}
\def\pareto#1#2{{#1}\otimes{#2}}
\def\sdif#1#2{{#1}\triangle{#2}}
\def\add#1{\Delta^+{#1}}
\def\del#1{\Delta^-{#1}}
\def\relpref#1#2{[{#1}]_{#2}}

\begin{document}

\title{Database Querying under Changing Preferences
\thanks{Research supported by NSF grant IIS-0307434.
This paper is an expanded version of
\cite{ChFOIKS06}.}
}


\author{Jan Chomicki}

\institute{Dept. of Computer Science and Engineering\\ University
at Buffalo\\ Buffalo, NY 14260-2000\\ {\tt
chomicki@cse.buffalo.edu}}


\institute{Jan Chomicki \at 
Dept. of Computer Science and Engineering, University
at Buffalo, Buffalo, NY 14260-2000\\
              Tel.: +716-645-3180 x103\\
              Fax: +716-645-3464\\
              \email{chomicki@cse.buffalo.edu}           %  \\
}

\date{Received: date / Accepted: date}

\maketitle

\begin{abstract}
We present here a formal foundation for an iterative and incremental
approach to constructing and evaluating preference queries. 
Our main focus is on {\em query modification}: a query transformation approach which works by revising the 
preference relation in the query. We provide a detailed analysis of the cases
where the order-theoretic properties of the preference relation are preserved by the
revision. We consider a number of different revision operators: union,
prioritized and Pareto composition.
We also formulate algebraic laws that enable incremental evaluation of preference
queries.
Finally, we consider two variations of the basic
framework: finite restrictions of preference relations 
and weak-order extensions of strict partial order preference relations.
%Insert your abstract here. Include keywords, PACS and mathematical
%subject classification numbers as needed.
\end{abstract}

\section{Introduction}\label{s:intro}
The notion of {\em preference} is common in various contexts involving 
decision or choice. Classical utility theory \cite{Fish70} views preferences
as {\em binary relations}. This view has recently been
adopted in database research \cite{ChEDBT02,ChTODS03,Kie02,KiKo02}, where preference relations
are used in formulating {\em preference queries}. In AI, various approaches
to compact specification of preferences have been explored \cite{BouetalJAIR04}. The semantics
underlying such approaches typically relies on preference relations between worlds.

Preferences can be embedded into database query languages in several different ways.
First, \cite{ChEDBT02,ChTODS03,Kie02,KiKo02} propose to introduce a special operator {\em ``find all the
most preferred  tuples according to a given preference relation.''}
This operator is called {\em winnow} in \cite{ChEDBT02,ChTODS03}. A special case of winnow is called
{\em skyline} \cite{BoKoSt01} and has been recently extensively studied \cite{PaTaFuSe03,BalEDBT04}.
Second, \cite{AgWi00,HrPa04} assume that preference relations are defined using numeric utility functions
and queries return tuples ordered by the values of a supplied utility function.
It is well-known that numeric utility functions cannot represent all strict partial
orders \cite{Fish70}, not even those that occur in database applications in a natural
way \cite{ChTODS03}. For example, utility functions cannot capture skylines.
Also, ordered relations go beyond the classical relational model of data.
The evaluation and optimization of queries over such relations requires significant
changes to relational query processors and optimizers \cite{IlSIGMOD04}. On the other hand,
winnow can be seamlessly combined with any relational operators.

We adopt here the first approach, based on winnow, 
within the preference query framework of \cite{ChTODS03} (a similar model
was described in \cite{Kie02}). In this framework, preference relations between tuples are defined by
first-order logical formulas. 

\begin{example}\label{ex:car}
Consider the relation $Car(Make,Year)$
and the following preference relation $\succ_{C_1}$ between {\em Car} tuples:
\begin{quote}
{\em within each make, prefer a more recent car,}
\end{quote}
which can be defined as follows:
\[(m,y)\succ_{C_1}(m',y') \equiv m=m'\wedge y>y'.\]
The winnow operator $\wnnw{C_1}$ returns for every make the most recent car available.
Consider the instance $r_1$ of $Car$ in Figure \ref{fig:winnow}a.
The set of tuples $\wnnw{C_1}(r_1)$ is shown in Figure \ref{fig:winnow}b.

%\begin{tabular}{ll}
%\begin{minipage}[t]{2in}
\begin{figure}[htb]
\centering
\mbox{%
\subfigure[]{%
\begin{tabular}{|l|l|l|}
\hline
&{\em Make} &{\em Year} \\\hline
$t_1$& VW & 2002\\
$t_2$& VW& 1997 \\
$t_3$& Kia & 1997\\\hline
\end{tabular}}\hspace{1.5in}
%\caption{The Car relation}
%\label{fig:car}
%\end{figure}
%\end{minipage}
%&
%\begin{minipage}[t]{2in}
%\begin{figure}[htb]
%\centering
\subfigure[]{%
\begin{tabular}{|l|l|l|}
\hline
&{\em Make} &{\em Year} \\\hline
$t_1$& VW & 2002\\
$t_3$& Kia & 1997\\\hline
\end{tabular}}}
\caption{(a) The Car relation; (b) Winnow result}
\label{fig:winnow}
\end{figure}
%\end{minipage}
%\end{tabular}
\end{example}

In this paper, we focus on preference queries of the form $\wnnw{\succ}(R)$, consisting of a single occurrence of winnow.
Here $\succ$ is a preference relation (typically defined by a formula), and $R$ is a database relation.
The relation $R$ represents the space of possible choices.
We also briefly  discuss how our results can be applied to more general
preference queries.

Past work on preference queries has made the assumption that preferences are {\em static}.
However, this assumption is often not satisfied. User preferences change, sometimes 
as a direct consequence of evaluating a preference query.
Therefore, we view preference querying as a {\em dynamic, iterative process}. The user
submits a query and inspects the result. The result may be satisfactory,
in which case the querying process terminates. Or, the result may be too large
or too small, contain unexpected answers, or fail to contain expected answers.
By inspecting the query answer, the user may realize some previously
unnoticed aspects of her preferences. 
It is also possible that not all the relevant data was included in the database
over which the preference query is evaluated.

So if the user is not satisfied with the preference query result, she has several further options:

{\em Modify and resubmit} the query. This is appropriate 
if the user decides to refine or change her preferences. For example, the
user may have started with a partial or vague concept of her preferences \cite{PuFaTo03}.
We consider here query modification consisting of {\em revising} the preference
relation $\succ$, although, of course, more general transformations may also be envisioned.

{\em Update} the database. This is appropriate if the user
discovers that there are more (or fewer) possible choices than originally
envisioned. For example, in comparison shopping the user may have discovered
a new source of relevant data.

In this context we pursue the following research challenges:

{\em Defining a repertoire of suitable preference relation revisions.}
In this work, we consider revisions obtained by {\em composing} the
original preference relation with a new preference relation, and
{\em transitively closing} the result (to guarantee transitivity).
We study different composition operators: union, and prioritized
and Pareto composition. Those operators represent several basic ways of combining preferences
and have already been incorporated into 
preference query languages \cite{ChTODS03,Kie02}.
The operators reflect different user attitudes
towards {\em preference conflicts}. (A conflict is, intuitively, a situation in which two preference
relations order the same pair of tuples differently.) Union ignores conflicts (and thus
such conflicts need to be prevented if we want to obtain a preference
relation which is a strict partial order). Prioritized composition
resolves preference conflicts by consistently giving priority to one
of the preference relations. Pareto composition resolves conflicts in a symmetric way.
We emphasize that revision is done using composition because we want the revised preference
relation to be uniquely defined  in the same first-order language as the original preference relation.
Clearly, the revision repertoire that we study in this paper does not exhaust
all meaningful scenarios. One can also imagine approaches where axiomatic properties
of preference revisions are studied, as in belief revision \cite{GaRo95}.

{\em Identifying essential properties of preference revisions.}  We
claim that revisions should preserve the order-theoretic properties of
the original preference relations. For example, if we start with a
preference relation which is a strict partial order, the revised
relation should also have those properties. This motivates, among others, transitively
closing  preference relations to guarantee transitivity. Preserving
order-theoretic properties of preference relations is particularly
important in view of the iterative construction of preference
queries where the output of a revision can serve as the input to another one.
We study both necessary and sufficient conditions on the original and revising
preference relations that yield the preservation of their 
order-theoretic properties. Necessary conditions
are connected with the absence of preference conflicts.
However, such conditions are typically not sufficient and stronger assumptions about
the preference relations need to be made. Somewhat surprisingly, a special class
of strict partial orders, {\em interval orders}, plays an important role in this context.
The conditional preservation results
we establish in this paper supplement those in \cite{ChTODS03,Kie02} and may
be used in other contexts where preference relations are composed,
for example in the implementation of preference query languages.
Another desirable property of revisions
is minimality in some well-defined sense. We define minimality in terms of symmetric
difference of preference relations but there are clearly other possibilities.

{\em Incremental evaluation of preference queries.} At each point of the
interaction with the user, the results of evaluating {\em previous}
versions of the given preference query are available. Therefore, they can be used to make
the evaluation of the {\em current} query more efficient. 
For both the preference revision and database update scenarios,
we formulate
algebraic laws that validate new query evaluation plans that use
materialized results of past query evaluations.
The laws use order-theoretic properties of preference relations in an essential way.

\begin{example}\label{ex:car:1}
Consider Example \ref{ex:car}.
Seeing the result of the query $\wnnw{C_1}(r_1)$, a user may realize that the preference 
relation $\succ_{C_1}$ is not quite what she had in mind. The result of the query
may contain some unexpected or unwanted tuples, for example $t_3$.
Thus the preference relation needs to be modified, for example
by revising it with the following preference relation $\succ_{C_2}$:
\[(m,y)\succ_{C_2}(m',y') \equiv m={\rm ''VW''}\wedge m'\not={\rm ''VW''}\wedge y=y'.\]
As there are no conflicts between $\succ_{C_1}$ and $\succ_{C_2}$,
the user chooses union as the composition operator. However, to
guarantee transitivity of the resulting preference relation, $\succ_{C_1}\cup\succ_{C_2}$ has to be
transitively closed. So the revised relation is $\succ_{C*}\equiv TC(\succ_{C_1}\cup\succ_{C_2}$).
(The explicit definition of $\succ_{C*}$ is given in Example \ref{ex:car:2}.)
The tuple $t_3$ is now dominated by $t_2$ (i.e., $t_2\succ_{C*} t_3$) and will not be returned to the user.
\end{example}

The plan of the paper is as follows.
In Section \ref{sec:basic}, we define the basic notions.
In Section \ref{sec:revision}, we introduce preference revision.
In Section \ref{sec:preserve}, we discuss query modification and
the preservation by revisions  of order-theoretic properties of preference relations.
In Section \ref{sec:incr}, we discuss incremental evaluation of preference
queries in the context of query modification and database updates.
Subsequently, we consider two variations of our basic
framework: (finite) restrictions of preference relations (Section  \ref{sec:finite})
and weak-order extensions of strict partial order preference relations
(Section \ref{sec:ext}).
We briefly discuss related work in Section \ref{sec:related}
and conclude in Section \ref{sec:concl}.

\section{Basic notions}\label{sec:basic}
We are working in the context of the relational model of data.
Relation schemas consist of finite sets of attributes.
For concreteness, we consider two infinite, countable domains: $\cald$ (uninterpreted constants, for readability shown as strings) and $\calq$ (rational numbers), but our results, except where explicitly indicated,
hold also for finite domains.
%Other domains could be considered as well without influencing most of the results of the paper.
%The result of a query $Q$ in a database instance $r$ is denoted $Q(r)$.
We assume that database instances are finite sets of tuples.
Additionally,
we have the standard built-in predicates.

\subsection{Preference relations}

We adopt here the framework of \cite{ChTODS03}.

\begin{definition}\label{def:prefrel}
Given a relation schema $R(A_1 \cdots A_k)$
such that $U_i$, $1\leq i\leq k$, is the domain (either $\cald$ or $\calq$)
of the attribute $A_i$, a relation $\succ$ is a {\em preference relation over $R$}
if it is a subset of $(U_1\times\cdots\times U_k)\times (U_1\times\cdots\times U_k)$.
\end{definition}

Although we assume that database instances are finite, in the presence of
infinite domains preference relations can be infinite.

%Intuitively, $\succ$ is a binary relation between tuples from the
%same (database) relation.
%We say that a tuple $t_1$ {\em dominates} a tuple $t_2$
%in $\succ$ if $t_1\succ t_2$.

Typical properties of a preference relation $\succ$ include \cite{Fish70}:
\begin{itemize}
\item {\em irreflexivity}: $\forall x.\ x\not\succ x;$
%\item {\em asymmetry}: $\forall x,y.\ x\succ y\Rightarrow y\not\succ x,$
\item {\em transitivity}: $\forall x,y,z.\ (x\succ y \wedge y\succ z)\Rightarrow x\succ z;$
\item {\em negative transitivity}: $\forall x,y,z.\ (x\not\succ y \wedge y\not\succ z)\Rightarrow x\not\succ z;$
\item {\em connectivity}: $\forall x,y.\ x\succ y\vee y\succ x \vee x=y;$
\item {\em strict partial order} (SPO) if $\succ$  is
irreflexive and transitive;
\item {\em interval order} (IO) \cite{Fish85} if $\succ$ is an SPO
and satisfies the condition \[\forall x,y,z,w.\ (x\succ y \wedge z\succ w)\Rightarrow (x\succ w \vee z\succ y);\]
\item {\em weak order} (WO) if $\succ$ is a
negatively transitive SPO;
\item {\em total order} if $\succ$ is
a connected SPO.
\end{itemize}
Every total order is a WO; every WO is an IO.
%; every IO is an SPO.

\begin{definition}\label{def:prefformula}
A {\em preference formula (pf)} $C(t_1,t_2)$ is a first-order
formula defining a preference relation $\succ_C$ in the standard
sense, namely
\[t_1\succ_C t_2\;{\rm iff}\; C(t_1,t_2).\]
An {\em intrinsic preference formula (ipf)} is a preference formula that
uses only built-in predicates.
\end{definition}

By using the notation $\succ_C$ for a preference relation, we assume that there is
an underlying pf $C$.
Occasionally, we will limit our attention
to ipfs consisting of the following two kinds of atomic formulas
(assuming we have two kinds of variables: $\cald$-variables and $\calq$-variables):
\begin{itemize}
\item {\em equality  constraints}: $x=y$, $x\not=y$, $x=c$, or $x\not= c$, where
$x$ and $y$ are $\cald$-variables, and $c$ is an uninterpreted constant;
\item {\em rational-order constraints}: $x\lambda y$ or $x\lambda c$, where
\mbox{$\lambda\in\{=,\not=,<,>,\leq,\geq\}$,} $x$ and $y$ are $\calq$-variables, and $c$ is a
rational number.
\end{itemize}
An ipf all of whose atomic formulas are equality  (resp. rational-order)
constraints will be called an {\em equality} (resp. {\em rational-order}) ipf.
If both equality and rational-order constraints are allowed in a formula, the
formula will be called {\em ERO}.
Clearly, ipfs are a special case of general constraints
\cite{CDB00,KaKuRe95}, and define {\em fixed}, although possibly infinite,
relations.

\begin{proposition}\label{prop:NP}
Satisfiability of quantifier-free ERO formulas 
is in NP.
\end{proposition}
\begin{proof}
Satisfiability of conjunctions of atomic ERO constraints
can be checked in linear time \cite{GuSuWe96}.
In an arbitrary quantifier-free ERO formula negation can be eliminated.
Then in every disjunction one needs to nondeterministically select one disjunct,
ultimately obtaining a conjunction of atomic constraints.\qed

\end{proof}

Proposition \ref{prop:NP} implies that all the properties that can be
polynomially reduced to validity of ERO formulas,
for example all the order-theoretic properties listed above, can be decided
in co-NP.

Every preference relation $\succ$ generates an indifference relation $\sim$:
two tuples $t_1$ and $t_2$ are {\em indifferent}
($t_1\sim t_2$) if
neither is preferred to the other one, i.e.,
$t_1\not\succ t_2$ and $t_2\not\succ t_1$.
We  denote by $\sim_C$ the indifference relation generated by $\succ_C$.

Composite preference relations are defined from simpler ones
using logical connectives. We focus on the following basic ways of composing preference
relations over the same schema: 
\begin{itemize}
\item {\em union:} 
$t_1\ (\union{\succ_1}{\succ_2})\ t_2\ {\rm iff}\ t_1\succ_1 t_2\vee t_1\succ_2 t_2;$
\item {\em prioritized composition:}
% (where $\sim_1$ is the indifference relation generated by $\succ_1$): 
$t_1\ (\prior{\succ_1}{\succ_2})\ t_2\ {\rm iff}\  t_1\succ_1 t_2 \vee (t_2\not\succ_1 t_1\wedge t_1\succ_2 t_2);$
\item {\em Pareto composition:}
\[t_1\ (\pareto{\succ_1}{\succ_2})\ t_2\ {\rm iff}\ (t_1\succ_1 t_2 \wedge t_2\not\succ_2 t_1) \vee 
(t_1\succ_2 t_2 \wedge t_2\not\succ_1 t_1).\]
\end{itemize}
We will use the above composition operators to construct revisions of given preference relations.
We also consider transitive closure:
\begin{definition}\label{def:transitive}
The {\em transitive closure} of a  preference relation
$\succ$ over a relation schema $R$ is a  preference relation 
$TC(\succ)$ over $R$ defined as: 
\[(t_1,t_2)\in TC(\succ)\;{\rm iff}\; t_1\succ^n t_2\;
{\rm for\; some\;} n> 0,\]
%\bigvee_{n\geq 1}\Phi^n(t_1,t_2)\] 
where:
\[\begin{array}{l}
t_1\succ^1 t_2\equiv t_1\succ t_2\\
t_1\succ^{n+1}t_2\equiv\exists t_3.\ 
t_1\succ t_3\wedge t_3\succ^n t_2.\\
%\Phi^1(t_1,t_2)\equiv\Phi(t_1,t_2)\\
%\Phi^{n+1}(t_1,t_2)\equiv \exists\bar{z}.
%\Phi(t_1,\bar{z})\wedge \Phi^n(\bar{z},t_2).\\
\end{array}\]
\end{definition}

Clearly, in general Definition \ref{def:transitive} leads to infinite formulas.
However, in the cases that we consider in this paper the preference
relation $\succ_{\tc{\succ}}$ will in fact be defined by a finite formula.

\begin{proposition}\label{prop:terminate}
Transitive closure of every preference relation defined by an ERO ipf
is definable using an ERO ipf of at most exponential size, which can be computed in exponential time.
\end{proposition}
\begin{proof}
This is because transitive closure can be expressed in Datalog and the 
evaluation of Datalog programs over equality and rational-order constraints
terminates in exponential  time (combined complexity) \cite{KaKuRe95}.\qed
\end{proof}

In the cases mentioned above, the transitive
closure of a given preference relation is a  relation definable in the signature of the preference formula.
But clearly transitive closure, unlike union and prioritized or Pareto composition,
is itself not  a first-order definable operator.

\subsection{Winnow}
We define now an algebraic operator that picks from a given relation the
set of the {\em most preferred tuples}, according to a given preference relation.
\begin{definition}\label{def:winnow}{\rm \cite{ChTODS03}}
If $R$ is a relation schema and $\succ$ a preference
relation over $R$,
then the {\em winnow operator} is written as $\wnnw{\succ}(R)$,
and for every instance $r$ of $R$:
\[\wnnw{\succ}(r)=\{t\in r\mid\neg \exists t'\in r.\ t'\succ t\}.\]
\end{definition}
If a preference relation is defined using a pf $C$, we write simply $\wnnw{C}$ instead of $\wnnw{\succ_C}$.
A {\em preference query} is a relational algebra query containing at least
one occurrence of the winnow operator.

\section{Preference revisions}\label{sec:revision}

%In the following discussion, it is not necessary to assume that preference relations are defined
%using logic formulas. Therefore, we talk about arbitrary preference relations, usually satisfying
%properties of partial and weak orders.

The basic setting is as follows: We have an {\em original} preference relation $\succ$ and revise it with a 
{\em revising\/} preference relation $\succ_0$ to obtain a {\em revised\/} preference relation $\succ'$.
We also call $\succ'$ a {\em revision\/} of $\succ$. 
We assume that $\succ$, $\succ_0$, and $\succ'$ are preference relations over the same schema,
and that all of them satisfy at least the properties of SPOs.

In our setting, a revision is obtained by
composing $\succ$ with $\succ_0$ using union, prioritized or Pareto composition, and transitively closing the result 
(if necessary to obtain transitivity).
However, we formulate some properties, like minimality or compatibility, in more general terms.

To define minimality, we order revisions using the symmetric difference ($\triangle$).

\begin{definition}\label{def:closeness}
Assume $\succ_1$ and $\succ_2$ are two revisions of  a preference relation $\succ$ 
with a preference relation $\succ_0$. We say that $\succ_1$ is {\em closer} than $\succ_2$  to $\succ$
if $\sdif{\succ_1}{\succ}\, \subset\, \sdif{\succ_2}{\succ}$.
\end{definition}

%For finite domains and SPOs, the closeness order defined above concides with the order
%based on the partial-order distance \cite{Bog73} of the revision to the original relation $\,\succ$.
\ignore{
\begin{definition}
A {\em minimal (resp. least)} revision of $\;\succ$ with $\succ_0$ is a revision 
that is minimal 
(resp. least) in the closeness
order among all revisions of $\;\succ$ with $\succ_0$.
\end{definition}
}

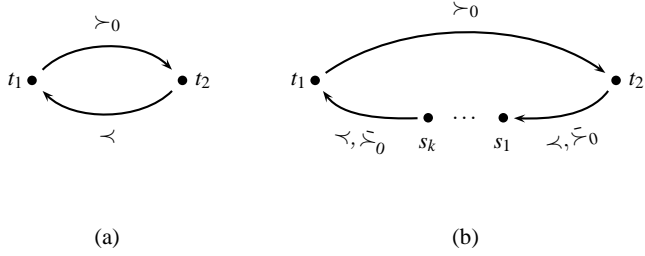
\begin{figure}[h]
\centering
\mbox{%
\subfigure[]{%
\begin{pspicture}(0,-1.5)(3,1.5)
\rput(.5,0){\rnode{T1}{\psdots(0,0)}}
\rput[r](.5,0){\rput(-6pt,0){$t_1$}}
\rput(2.5,0){\rnode{T2}{\psdots(0,0)}}
\rput[l](2.5,0){\rput(8pt,0){$t_2$}}
\psset{nodesep=4pt}
\ncarc[arcangleA=45,arcangleB=45]{->}{T1}{T2}
\aput{:U}{$\succ_0$}
\ncarc[arcangleA=-45,arcangleB=-45]{<-}{T1}{T2}
\bput{:U}{$\prec$}
\end{pspicture}
}
%\caption{$0$-conflict}
%\label{0-conflict}
%\end{figure}
\quad\quad
\subfigure[]{%
%\begin{figure}[h]
\begin{pspicture}(0,-1.5)(5,2)
\rput(.5,0){\rnode{T1}{\psdots(0,0)}}
\rput[r](.5,0){\rput(-6pt,0){$t_1$}}
\rput(4.5,0){\rnode{T2}{\psdots(0,0)}}
\rput[l](4.5,0){\rput(8pt,0){$t_2$}}
\rput(2,-.5){\rnode{SK}{\psdots(0,0)}}
\rput[t](2,-.5){\rput(0,-10pt){$s_k$}}
\rput(3,-.5){\rnode{S1}{\psdots(0,0)}}
\rput[t](3,-.5){\rput(0,-10pt){$s_1$}}
\rput(2.5,-.5){$\ldots{}$}
\psset{nodesep=4pt}
\ncarc[arcangleA=35,arcangleB=35]{->}{T1}{T2}
\aput{:U}{$\succ_0$}
\ncarc[arcangleA=-35,arcangleB=-20]{<-}{T1}{SK}
\bput{:U}{$\prec,\bar{\succ}_0$}
\ncarc[arcangleA=-20,arcangleB=-35]{<-}{S1}{T2}
\bput{:U}{$\prec,\bar{\succ}_0$}
\end{pspicture}
}}
\caption{(a) $0$-conflict; (b) $1$-conflict}
\label{1-conflict}
\end{figure}

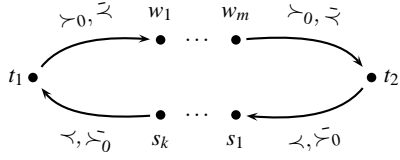
\begin{figure}[h]
\begin{center}
\begin{pspicture}(0,-2)(6,2)
\rput(.5,0){\rnode{T1}{\psdots(0,0)}}
\rput[r](.5,0){\rput(-6pt,0){$t_1$}}
\rput(5,0){\rnode{T2}{\psdots(0,0)}}
\rput[l](5,0){\rput(8pt,0){$t_2$}}
\rput(2.2,.5){\rnode{W1}{\psdots(0,0)}}
\rput[b](2.2,.5){\rput(0,10pt){$w_1$}}
\rput(3.2,.5){\rnode{WM}{\psdots(0,0)}}
\rput[b](3.2,.5){\rput(0,10pt){$w_m$}}
\rput(2.7,.5){$\ldots{}$}
\rput(2.2,-.5){\rnode{SK}{\psdots(0,0)}}
\rput[t](2.2,-.5){\rput(0,-10pt){$s_k$}}
\rput(3.2,-.5){\rnode{S1}{\psdots(0,0)}}
\rput[t](3.2,-.5){\rput(0,-10pt){$s_1$}}
\rput(2.7,-.5){$\ldots{}$}

\psset{nodesep=4pt}
\ncarc[arcangleA=35,arcangleB=20]{->}{T1}{W1}
\aput{:U}{$\succ_0,\bar{\prec}$}
\ncarc[arcangleA=20,arcangleB=35]{->}{WM}{T2}
\aput{:U}{$\succ_0,\bar{\prec}$}
\ncarc[arcangleA=-35,arcangleB=-20]{<-}{T1}{SK}
\bput{:U}{$\prec,\bar{\succ_0}$}
\ncarc[arcangleA=-20,arcangleB=-35]{<-}{S1}{T2}
\bput{:U}{$\prec,\bar{\succ_0}$}
\end{pspicture}
\end{center}
\caption{$2$-conflict}
\label{2-conflict}
\end{figure}

To further describe the behavior of revisions, we define several kinds of {\em preference conflicts}.
The intuition here is to characterize those conflicts that, when eliminated by 
prioritized or Pareto composition, reappear if the resulting preference relation is closed
by transitivity.
\begin{definition}
A {\em $0$-conflict} between a preference relation $\;\succ$  and a preference relation $\;\succ_0$ 
is a pair $(t_1,t_2)$ such that \mbox{$t_1\succ_0 t_2$} and \mbox{$t_2\succ t_1$}.
A {\em $1$-conflict} between $\;\succ$  and $\;\succ_0$ is a pair $(t_1,t_2)$ such that $t_1\succ_0 t_2$ 
and there exist $s_1,\ldots s_k$, $k\geq 1$, such that
$t_2\succ s_1\succ\cdots\succ s_k\succ t_1$ and $t_1\not\succ_0 s_k \not\succ_0\cdots\not\succ_0 s_1\not\succ_0 t_2$.
A {\em $2$-conflict} between $\;\succ$  and $\;\succ_0$ is a pair $(t_1,t_2)$ such that 
there exist $s_1,\ldots, s_k$, $k\geq 1$ and $w_1,\ldots,w_m$, $m\geq 1$, such that
$t_2\succ s_1\succ\cdots\succ s_k\succ t_1$, $t_1\not\succ_0 s_k \not\succ_0\cdots\not\succ_0 s_1\not\succ_0 t_2$,
$t_1\succ_0 w_1\succ_0\cdots\succ_0 w_m\succ t_2$, and $t_2\not\succ w_m \not\succ\cdots\not\succ w_1\not\succ t_1$
\end{definition}
A $1$-conflict is a $0$-conflict if $\succ$ is an SPO, but not necessarily vice versa.
A $2$-conflict is a $1$-conflict if $\succ_0$ is an SPO.
The different kinds of conflicts are pictured in Figures \ref{1-conflict} and \ref{2-conflict}
($\bar{\succ}$ denotes the complement of $\succ$).
\begin{example}
If $\succ_0=\{(a,b)\}$ and $\succ=\{(b,a)\}$, then $(a,b)$ is a $0$-conflict which is not a $1$-conflict. 
If we add $(b,c)$ and $(c,a)$ to $\succ$, then the conflict becomes a $1$-conflict ($s_1=c$).
If we further add $(c,b)$ or $(a,c)$ to $\succ_0$, then the conflict is not a $1$-conflict anymore.
On the other hand, if we add $(a,d)$ and $(d,b)$ to $\succ_0$ instead, then we obtain a $2$-conflict.
\end{example}

We assume here that the preference relations $\succ$ and $\succ_0$ are SPOs.
If $\succ'=\tc{\union{\succ}{\succ_0}}$, then for every $0$-conflict between $\succ$ and $\succ_0$, we still
obviously have $t_1\succ' t_2$ and $t_2\succ' t_1$. Therefore, we say that the union does not resolve
any conflicts.
On the other hand, if $\succ'=\tc{\prior{\succ_0}{\succ}}$, then for each $0$-conflict $(t_1,t_2)$,
$t_1\ \prior{\succ_0}{\succ}\ t_2$ and $\neg(t_2\ \prior{\succ_0}{\succ}\ t_1)$.
In the case of $1$-conflicts, we get again $t_1\succ' t_2$ and $t_2\succ' t_1$.
But in the case where a $0$-conflict is not a $1$-conflict, we get only $t_1\succ' t_2$.
Thus we say that prioritized composition resolves  those $0$-conflicts
that are not $1$-conflicts.
Finally, if $\succ'=\tc{\pareto{\succ}{\succ_0}}$, then for each $1$-conflict $(t_1,t_2)$,
$\neg(t_1\ \pareto{\succ}{\succ_0}\ t_2)$ and $\neg(t_2\ \pareto{\succ}{\succ_0}\ t_1)$. 
We get $t_1\succ' t_2$ and $t_2\succ' t_1$ if the conflict is a $2$-conflict,
but if it is not, we obtain only $t_2\ \succ'\ t_1$.
Thus we say that Pareto composition resolves those $1$-conflicts that are not $2$-conflicts.
(Pareto composition resolves also conflicts that are symmetric versions of $1$-conflicts,
with $\succ_0$ and $\succ$ interchanged,  which are not $2$-conflicts.)

We now characterize those combinations of $\succ$  and $\succ_0$ that avoid different kinds of conflicts.
\begin{definition}\label{def:compat}
A preference relation $\succ$  is {\em $i$-compatible}($i=0,1,2$)  with a preference relation
$\succ_0$  if there are no $i$-conflicts between $\succ$  and $\succ_0$.
\end{definition}
$0$- and $2$-compatibility are symmetric. $1$-compatibility is not necessarily symmetric.
For SPOs, $0$-compatibility implies $1$-compatibility and $1$-compatibility implies $2$-compatibility.
Examples \ref{ex:car} and \ref{ex:car:1} show a pair
of $0$-compatible relations. 
$0$-compatibility of $\succ$ and $\succ_0$ {\em does not require} the acyclicity of $\succ\cup\succ_0$ or that one of the following hold:
$\succ\ \subseteq\ \succ_0$, $\succ_0\ \subseteq\ \succ$, or $\;\succ\cap\succ_0\ =\ \emptyset$.
%For the former, consider $\succ=\{(a,b),(c,d)\}$ and $\succ_0=\{(b,c),(d,a)\}$.
%For the latter, consider $\succ=\{(a,b),(b,c),(a,c)\}$ and $\succ_0=\{(a,b),(a,d)\}$.
% two SPOs which are compatible.

Propositions \ref{prop:NP} and \ref{prop:terminate} imply that
all the variants of compatibility defined above are decidable for
ERO ipfs.
For example, $1$-compatibility is expressed by the condition
$\;\succ_0^{-1}\,\cap\, \tc{\succ\! -\! \succ_0^{-1}}=\emptyset$ where $\succ_0^{-1}$ is the inverse
of the preference relation $\succ_0$.

$0$-compatibility of $\succ$ and $\succ_0$ is a {\em necessary} condition
for $\tc{\union{\succ}{\succ_0}}$ to be irreflexive, and thus an SPO.
Similar considerations apply to $\tc{\prior{\succ_0}{\succ}}$ and $1$-compatibility,
and $\tc{\pareto{\succ}{\succ_0}}$ and $2$-compatibility.
In the next section, we will see that those conditions are not {\em sufficient}: further restrictions on the
preference relations will be introduced.

We conclude by noting the relationships between the three notions
of preference composition introduced above.
\begin{lemma}\label{l:coincide}
For every preference relations $\succ$ and $\succ_0$
\[\pareto{\succ_0}{\succ}\subseteq \prior{\succ_0}{\succ}\subseteq \union{\succ_0}{\succ},\]
and if $\,\succ_0$ and $\succ$ are $0$-compatible
\[\pareto{\succ_0}{\succ}=\prior{\succ_0}{\succ}=\union{\succ_0}{\succ}.\]
\end{lemma}

\section{Query modification}\label{sec:preserve}
In this section, we study preference query modification
\footnote{The  term {\em query modification} was used in early relational systems
like INGRES to denote a technique that produced a changed version of a query
submitted by a user. The changes were meant to incorporate view definitions,
integrity constraints and security specifications. We feel that it is justified to use
the same term in the context of composition of a preference relation in a query
with some other preference relation, to produce a new query.}.
A given
preference query $\wnnw{\succ}(R)$ is transformed to the query
$\wnnw{\succ'}(R)$ where $\succ'$ is obtained by composing
the original preference relation
$\succ$ with the revising preference relation $\succ_0$, and transitively closing the result.
(The last step is clearly unnecessary if the obtained preference relation is already transitive.)
We want $\succ'$ to satisfy the same order-theoretic properties as $\succ$ and $\succ_0$, and 
to be minimally different from $\succ$.
To achieve those goals, we impose additional conditions on $\succ$ and $\succ_0$.

For every $\theta\in\{\cup,\rhd,\otimes\}$, we consider the order-theoretic properties
of the preference relation $\succ'\ =\ \succ_0\theta\succ$, or $\succ'\ =\ \tc{\succ_0\theta\succ}$ if
$\succ_0\theta\succ$ is not guaranteed to be transitive.
To ensure that this preference relation is an SPO, only irreflexivity has to be
guaranteed; for weak orders one has also to establish negative transitivity.

\subsection{Strict partial orders}

SPOs have several important properties from the user's point of view, and thus
their preservation is desirable. For instance,
all the preference relations defined in \cite{Kie02} and in the language Preference SQL \cite{KiKo02} are SPOs. 
Moreover, if $\succ$ is an SPO, then the winnow $\wnnw{\succ}(r)$ is nonempty 
if (a finite) $r$ is nonempty.
The fundamental  algorithms for computing winnow require that the preference relation
be an SPO \cite{ChTODS03}. 
Also, in that case incremental evaluation of preference
queries becomes possible (Proposition \ref{prop:contain} and Theorem \ref{th:delta}).

\begin{theorem}\label{th:spo:union}
For every $0$-compatible preference relations $\succ$ and $\succ_0$ such that
one is an interval order (IO) and the other an SPO,
the preference relation $\, \tc{\succ_0\theta\succ}$, where $\theta\in\{\cup,\rhd,\otimes\}$,
is an SPO.
If the IO is a WO, then 
$\, \tc{\succ_0\theta\succ}=\succ_0\theta\succ$.
\end{theorem}
\begin{proof} 
By Lemma \ref{l:coincide}, $0$-compatibility implies that 
$\union{\succ_0}{\succ}=\prior{\succ_0}{\succ}=\pareto{\succ_0}{\succ}$.
Thus, WLOG we consider only union.
Assume $\succ_0$ is an IO.
If $\, \tc{\union{\succ}{\succ_0}}$ is not irreflexive, then $\union{\succ}{\succ_0}$
has a cycle. Consider such cycle of minimum length.
It consists of edges that are alternately labeled $\succ_0$ (only) and $\succ$ (only).
(Otherwise the cycle can be shortened).
If there is more than one non-consecutive $\succ_0$-edge in the cycle, then $\succ_0$ being an IO
implies that the cycle can be shortened. So the cycle consists of two edges:
$t_1\succ_0 t_2$ and $t_2\succ t_1$. But this is a $0$-conflict violating
$0$-compatibility of $\succ$ and $\succ_0$.\qed
\end{proof}

It is easy to see that there is no preference relation which is an SPO, contains
$\succ\cup\succ_0$, and is closer (in the sense of Definition \ref{def:closeness})
to $\succ$ than $\, \tc{\union{\succ}{\succ_0}}$.

As can be seen from the above proof, the fact that one of the preference
relations is an interval order makes it possible to eliminate those paths
(and thus also cycles) in $\, \tc{\union{\succ}{\succ_0}}$ that interleave
$\succ$ and $\succ_0$ more than once.
In this way acyclicity reduces to the lack of $0$-conflicts.

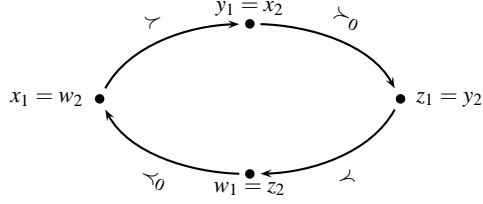
\begin{figure}
\begin{center}
\begin{pspicture}(-1.5,-2)(5.5,2)
\rput(0,0){\rnode{T1}{\psdots(0,0)}}
\rput[r](0,0){\rput[r](-6pt,0){$x_1=w_2$}}
\rput(2,1){\rnode{T2}{\psdots(0,0)}}
\rput[b](2,1.1){$y_1=x_2$}
\rput(4,0){\rnode{T3}{\psdots(0,0)}}
\rput[l](4,0){\rput[l](6pt,0){$z_1=y_2$}}
\rput(2,-1){\rnode{T4}{\psdots(0,0)}}
\rput[t](2,-1.1){$w_1=z_2$}

\psset{nodesep=4pt}
\ncarc[arcangleA=35,arcangleB=25]{->}{T1}{T2}
\aput{:U}{$\succ$}
\ncarc[arcangleA=25,arcangleB=35]{->}{T2}{T3}
\aput{:U}{$\succ_0$}
\ncarc[arcangleA=-25,arcangleB=-35]{<-}{T4}{T3}
\bput{:U}{$\prec$}
\ncarc[arcangleA=-35,arcangleB=-25]{<-}{T1}{T4}
\bput{:U}{$\prec_0$}
\end{pspicture}
\end{center}
\caption{A cycle for $0$-compatible relations that are not IOs.}
\label{fig:cycle}
\end{figure}

It seems that the interval order (IO) requirement in Theorem \ref{th:spo:union} cannot be weakened
without needing to strengthen the remaining assumptions.
If neither of $\,\succ$ and $\succ_0$ is an IO, then
we can find such elements $x_1$, $y_1$, $z_1$, $w_1$, $x_2$, $y_2$, $z_2$, $w_2$  that
\[x_1\succ y_1, z_1\succ w_1, x_1\not\succ w_1, z_1\not\succ y_1,
x_2\succ_0 y_2, z_2\succ_0 w_2, x_2\not\succ_0 w_2,\]    and $z_2\not\succ_0 y_2$.
If we choose $y_1=x_2$, $z_1=y_2$, $w_1=z_2$, and $x_1=w_2$, then we get a cycle in
$\union{\succ}{\succ_0}$. Note that in this case $\succ$ and $\succ_0$ are still $0$-compatible.
Also, there  is no SPO preference relation which contains $\union{\succ}{\succ_0}$
because each such relation has to contain $\, \tc{\union{\succ}{\succ_0}}$.
This situation is pictured in Figure \ref{fig:cycle}.

\begin{example}\label{ex:car:3}
Consider again the preference relation $\succ_{C_1}$:
\[(m,y)\succ_{C_1}(m',y') \equiv m=m'\wedge y>y'.\]
Suppose that the new preference information is captured as  $\succ_{C_3}$
which is an IO but not a WO:
\[\begin{array}{l}
(m,y)\succ_{C_3}(m',y') \equiv m={\rm ''VW''}\wedge y=1999\wedge m'={\rm ''Kia''}\wedge y'=1999.\end{array}\]
Then $\tc{\union{\succ_{C_1}}{\succ_{C_3}}}$, which properly contains 
$\union{\succ_{C_1}}{\succ_{C_3}}$, is defined as the SPO $\succ_{C_4}$:
\[\begin{array}{lcl}
(m,y)\succ_{C_4}(m',y')&\equiv &m=m'\wedge y>y'\vee \\
&&m={\rm ''VW''}\wedge y\geq 1999\wedge m'={\rm ''Kia''}\wedge y'\leq 1999.
\end{array}\]
%\[\begin{array}{l}
%(m,y)\succ_{C_4}(m',y')\equiv m=m'\wedge y>y'\\\vee\
% m={\rm ''VW''}\wedge y\geq 1999\wedge m'={\rm ''Kia''}\wedge y'\leq 1999.
%\end{array}\]

\end{example}

\ignore{
If the compatibility of $\succ$ and $\succ_0$ is relaxed, then it is easy
to obtain a cycle in $TC(\union{\succ}{\succ_0})$ and consequently, no strict partial
order refinement. One of the orders being an IO is also essential, as shown below.
\begin{theorem}\label{th:spo:greater}
There exist compatible preference relations $\succ$ and $\succ_0$ such that
(1) $\succ$ is an SPO and (2) $\succ_0$ is an SPO
of cardinality two, and 
there is no SPO refinement of $\succ$ with $\succ_0$.
\end{theorem}
\begin{proof}
Let a, b, c, and d tuples be in a database, $\succ=\{(a,b),(c,d)\}$ and
$\succ_{0}=\{(b,c),(d,a)\}$.
In this case, $TC(\union{\succ}{\succ_0})$  contains the tuple $(a,a)$ and thus is 
not irreflexive.
By Lemma \ref{lem:TC}, we know that  $TC(\union{\succ}{\succ_0})$ is 
contained in every transitive refinement. Therefore, every
transitive refinement is not irreflexive.
\end{proof}
}

Theorem \ref{th:spo:union} implies that if $\succ$ and $\succ_0$ are $0$-compatible and one of them
contains only one pair, then $TC(\union{\succ}{\succ_0})$ is an SPO.
So what will happen if we break up the preference
relation $\succ_{0}$ from Figure \ref{fig:cycle}
into two one-element relations $\succ_1$ and $\succ_2$ and attempt to 
apply Theorem \ref{th:spo:union} twice?
Unfortunately, such a ``strategy'' does not work. The second
step is not possible because the preference relation $\succ_2$
is not $0$-compatible with the revision of $\succ$ with $\succ_1$.

For dealing with {\em prioritized composition}, $0$-compatibility
can be replaced by a less restrictive condition, $1$-compatibility, because prioritized composition
already provides a way of resolving some conflicts. 
%In this way, nonmonotonicity of revisions becomes possible.

\begin{theorem}\label{th:spo:prior}
For every preference relations $\succ$ and $\succ_0$ such that
$\succ_0$ is an IO, $\succ$ is an SPO
and $\succ$ is $1$-compatible with $\succ_0$,
the preference relation $\tc{\prior{\succ_0}{\succ}}$ 
is an SPO.
\end{theorem}
\begin{proof} 
We assume that $\tc{\prior{\succ_0}{\succ}}$ is not irreflexive and  consider a cycle of minimum length 
in $\prior{\succ_0}{\succ}$.
If the  cycle has two non-consecutive edges labeled (not necessarily
exclusively) by $\succ_0$, then it can be shortened, because $\succ_0$ is an IO.
The cycle has to consist of an edge $t_1\succ_0 t_2$ and a sequence of edges
(labeled only by $\succ$):
$t_2\succ t_3,\ldots,t_{n-1}\succ t_n,t_n\succ t_1$ such that $n>2$.
and $t_1\not\succ_0 t_n\not\succ_0\ldots\not\succ_0 t_3\not\succ_0 t_2$.
(We cannot shorten sequences of consecutive $\succ$-edges because $\succ$
is not necessarily preserved in $\prior{\succ_0}{\succ}$.)
Thus $(t_1,t_2)$ is a $1$-conflict violating $1$-compatibility of $\succ$
with $\succ_0$.\qed
\end{proof}

Clearly, there is no SPO preference relation which contains
$\prior{\succ_0}{\succ}$, and is closer
to $\succ$ than $\, \tc{\prior{\succ_0}{\succ}}$.
Violating any of the conditions of Theorem \ref{th:spo:prior} may lead
to a situation in which no SPO preference relation which contains
$\prior{\succ_0}{\succ}$ exists. 
%To see that, construct a pair of preference
%relations that contain a $1$-conflict.

If $\succ_0$ is a WO, the requirement of $1$-compatibility and the computation of transitive closure are
unnecessary.
We first recall some basic properties of weak orders.
\begin{proposition}\label{prop:weak}
Let $\succ$ be a WO preference relation over a schema $R$ and $\sim$ the indifference
relation generated by $\succ$.
If $x\succ y$, $y\sim z$ and $z\succ w$, then also $x\succ z$ and $y\succ w$.
\end{proposition}

\begin{theorem}\label{th:spo:weak:priority}
For every preference relations $\succ_0$ and $\succ$ such that
$\succ_0$ is a WO and $\succ$ an SPO, the preference relation
$\prior{\succ_0}{\succ}$ is an SPO.
\end{theorem}
\begin{proof}
Clearly, $\succ'=\prior{\succ_0}{\succ}$, as a subset of $\union{\succ_0}{\succ}$, is irreflexive.
To show transitivity, consider $t_1 \succ' t_2$ and
$t_2 \succ' t_3$.
There are four possibilities:
(1) If $t_1\succ_0 t_2$ and $t_2 \succ_0 t_3$, then $t_1\succ_0 t_3$ and $t_1 \succ' t_3$.
(2) If $t_1\succ_0 t_2$, $t_3\not\succ_0 t_2$ and $t_2\succ t_3$,
then also $t_2\succ_0 t_3$ or $t_2\sim_0 t_3$ (where $\sim_0$ is the
indifference relation generated by $\succ_0$).
In either case, $t_1\succ_0 t_3$ and $t_1\succ' t_3$ (the second case requires using 
Proposition \ref{prop:weak}).
(3) $t_2\not\succ_0 t_1$, $t_1\succ t_2$ and $t_2\succ_0 t_3$: symmetric to (2).
(4) If $t_2\not\succ_0 t_1$, $t_1\succ t_2$, $t_3\not\succ_0 t_2$ and $t_2\succ t_3$,
then $t_3\not\succ_0 t_1$ (by the negative transitivity of $\succ_0$) and
$t_1\succ t_3$. Thus $t_1\succ' t_3$.\qed
\end{proof}

Let's turn now to {\em Pareto composition}.
There does not seem to be any simple way to {\em weaken} the
assumptions in Theorem \ref{th:spo:union} using the notion of $2$-compatibility.
Assuming that $\succ$, $\succ_0$,
or even both are IOs does not sufficiently restrict the possible interleavings
of $\succ$ and $\succ_0$ in $\tc{\pareto{\succ_0}{\succ}}$
because neither of those two preference relations is guaranteed to
be preserved in $\tc{\pareto{\succ_0}{\succ}}$.
However, we can establish a weaker version of Theorem \ref{th:spo:weak:priority}.

\begin{theorem}\label{th:spo:weak:pareto}
For every preference relations $\ \succ_0$ and $\succ$ such that
both are WOs, the preference relation
$\pareto{\succ_0}{\succ}$ is an SPO.
\end{theorem}
\begin{proof}
Similar to the proof of Theorem \ref{th:spo:weak:priority}.
\end{proof}

\ignore{
\begin{theorem}\label{th:spo:greater:priority}
There exist preference relations $\succ$ and $\succ_0$ such that
(1) $\succ$ is an SPO and (2) $\succ_0$ is an SPO,
and there is no SPO overriding revision of $\succ$ with $\succ_0$.
\end{theorem}
\begin{proof}
Consider a relation with three tuples $a$, $b$, and $c$.
Let $\succ_0=\{(b,a)\}$ and $\succ=\{(a,c),(c,b),(a,b)\}$.
Then there is a cycle in $\prior{\succ_0}{\succ}$.
\end{proof}
}
\ignore{
\begin{theorem}\label{th:spo:pareto}
For every preference relations $\succ$ and $\succ_0$ such that
$\succ_0$ and $\succ$ are IOs
and $\succ$ is $2$-compatible with $\succ_0$,
the preference relation $\tc{\pareto{\succ_0}{\succ}}$ 
is an SPO.
\end{theorem}
\begin{proof} 
Consider a minimum-length cycle in $\pareto{\succ_0}{\succ}$.
It has to contain at least one edge labelled exclusively by
$\succ_0$ and one edge labelled exclusively by $\succ$
(otherwise, one of those relations is not irreflexive).
Consider a maximal fragment $t_1,\ldots,t_n$ labelled with
$\succ_0$ of this cycle. Assume there is another edge
$t_3\succ_0 t_4$ in the cycle. Then $t_1\succ_0 t_4$ or
$t_3\succ_0 t_2$ (because $\succ_0$ is an IO). 
If $t_1 (\pareto{\succ_0}{\succ}) t_4$,
then the cycle can be shortened.
Otherwise, $t_4\succ t_1$. Then $t_4\succ t_3$
or $t_2 \succ t_1$ (because $\succ$ is an IO).
The first is impossible because $t_3 (\pareto{\succ_0}{\succ}) t_4$.
The second is also impossible because it leads to a $2$-conflict
\end{proof}
}

Proposition \ref{prop:terminate} implies that for all
preference relations defined using ERO ipfs,
the computation of the preference relations \mbox{$TC(\union{\succ}{\succ_0})$}, \mbox{$TC(\succ_0\rhd\succ)$}, 
as well as  \mbox{$TC(\pareto{\succ}{\succ_0})$} terminates.
The computation of transitive closure is done in a completely
database-independent way.
\begin{example}\label{ex:car:4}
Consider Examples \ref{ex:car} and  \ref{ex:car:3}. 
We can infer that
\[t_1=({\rm ''VW''},2002)\succ_{C_4} ({\rm ''Kia''},1997)=t_3,\]
because $({\rm ''VW''},2002)\succ_{C_1}({\rm ''VW''},1999),$
$({\rm ''VW''},1999)\succ_{C_3}({\rm ''Kia''},1999),$
and 
$({\rm ''Kia''},1999)\succ_{C_1}({\rm ''Kia''},1997).$
The tuples $({\rm ''VW''},1999)$ and $({\rm ''Kia''},1999)$ are {\em not\/} 
in the database.
\end{example}

If the conditions of Theorems \ref{th:spo:union} and \ref{th:spo:prior} do not apply,
Proposition \ref{prop:terminate} implies that for ERO ipfs
the computation of $\tc{\union{\succ}{\succ_0}}$,  \mbox{$TC(\succ_0\rhd\succ)$} 
and \mbox{$TC(\pareto{\succ}{\succ_0})$} yields
some finite ipf $C(t_1,t_2)$. Thus the irreflexivity of the resulting preference relation
reduces to the unsatisfiability of $\ C(t,t)$, which by Proposition \ref{prop:NP}
is a decidable problem for ERO ipfs. Of course, the relation, being a transitive closure, is already transitive.

\begin{example}\label{ex:car:2}
Consider Examples \ref{ex:car} and \ref{ex:car:1}.
Neither of the preference relations $\succ_{C_1}$ and $\succ_{C_2}$ is an interval order.
Therefore, the results established earlier in this section do not apply.
The preference relation $\succ_{C*}=\tc{\union{\succ_{C_1}}{\succ_{C_2}}}$ is defined
as follows (this definition is obtained using Constraint Datalog computation):
\[\begin{array}{lcl}
(m,y)\succ_{C*}(m',y') &\equiv & m=m'\wedge y>y'\vee\\
&& m={\rm ''VW''}\wedge m'\not={\rm ''VW''}\wedge y\geq y'.\\ 
\end{array}\]
%\[\begin{array}{l}
%(m,y)\succ_{C*}(m',y') \ \equiv\ m=m'\wedge y>y'\hspace{50pt}\\ 
%\hfill\vee\ m={\rm ''VW''}\wedge m'\not={\rm ''VW''}\wedge y\geq y'\\ 
%\end{array}\]

The preference relation $\succ_{C*}$ is irreflexive (this can be effectively checked).
It also properly contains $\union{\succ_{C_1}}{\succ_{C_2}}$, because
$t_1\succ_{C*} t_3$ but $t_1\not \succ_{C_1}t_3$ and $t_1\not \succ_{C_2}t_3$.
The query $\wnnw{C*}(Car)$ evaluated in the instance $r_1$ (Figure \ref{fig:winnow}) returns only the tuple $t_1$. 
\end{example}

\ignore{
To determine whether a specific combination of $\succ$ and $\succ_0$
has the least WO refinement (resp. least WO overriding revision), 
one can check whether $TC(\union{\succ}{\succ_0})$ (resp. $TC(\prior{\succ_0}{\succ})$) is 
irreflexive and negatively transitive.
}

\subsection{Weak orders}

\ignore{
We start by establishing a number of auxiliary results about weak orders.
By $\sim$ we denote the indifference relation generated by a preference relation $\succ$.
\begin{lemma}\label{lem:weak:1}
For every weak order preference relation $\succ$, and every x,y,
and z: if x$\sim$y and either x$\succ$z or y$\succ$z, then
x$\succ$z and y$\succ$z.
\end{lemma}
\begin{lemma}\label{lem:weak:2}
For every weak order preference relation $\succ$,
and every x,y,and
z: if x$\sim$y and either x$\not$$\succ$z or 
y$\not$$\succ$z, then x$\not$$\succ$z and y$\not$$\succ$z.
\end{lemma}
}
Weak orders are practically important because they capture the situation where
the domain can be decomposed into layers such that the layers are totally ordered and all the elements in one layer are
mutually indifferent. This is the case, for example, if a preference relation can be represented using a numeric
utility function.
If a preference relation is a WO, a particularly efficient (essentially single pass)
algorithm for computing winnow is applicable \cite{ChCDB04}.

We will see that for weak orders the transitive closure computation is unnecessary
and minimal revisions are directly definable in terms of the preference relations involved.
\ignore{
We first consider combinations of SPOs and weak orders.
\begin{theorem}\label{th:spo:weak}
For every $0$-compatible preference relations $\succ$ and $\succ_0$ such that
one is an SPO and the other
a WO, the preference relation
$\union{\succ}{\succ_{0}}$ is an SPO.
\end{theorem}
\begin{proof}
The relation $\union{\succ}{\succ_0}$ is irreflexive because  both
$\succ$ and $\succ_0$ are irreflexive.
Transitivity is proved by case analysis using Lemma \ref{lem:weak:1}.
\end{proof}
Clearly, the refinement yielding $\union{\succ}{\succ_0}$ under the assumptions of
Theorem \ref{th:spo:weak} cannot always be a WO.
As a counterexample, consider for $\succ$ any SPO which is not a WO,
for example:
\[a\succ b, a\sim c, b\sim c,\]
and for $\succ_0$ an empty order. 
Notice that in the above example the preference relation $\succ$ has  multiple minimal WO 
extensions which are also minimal WO refinements. Therefore, there
is no least WO refinement.
}

\begin{theorem}\label{th:weak:weak}
For every $0$-compatible WO preference relations $\succ$ and $\succ_0$,
the preference relations $\union{\succ}{\succ_{0}}$ and $\pareto{\succ}{\succ_{0}}$ are WO.
\end{theorem}
\begin{proof}
In view of Lemma \ref{l:coincide}, we can consider only $\succ'=\union{\succ}{\succ_{0}}$. 

Irreflexivity is obvious.
For transitivity, assume $t_1\succ' t_2$ and $t_2\succ' t_3$.
If $t_1\succ t_2\succ t_3$ (resp. $t_1\succ_0 t_2\succ_0 t_3$),
then $t_1\succ t_3$ (resp. $t_1\succ_0 t_3)$ and $t_1\succ' t_3$.
If $t_1\succ_0 t_2$ and $t_2\succ t_3$, we need $0$-compatibility
to infer that $t_2\not\succ t_1$ and thus $t_1\succ t_2$ or
$t_1\sim t_2$ (where $\sim$ is the indifference relation generated by
$\succ$). In both cases, we can infer $t_1\succ t_3$ and thus
$t_1\succ' t_3$. The last case is symmetric to the previous one.

For negative transitivity, consider $t_1\not\succ' t_2$ and $t_2\not\succ' t_3$.
Then $t_1\not\succ_0 t_2$, $t_2\not\succ_0 t_3$,
$t_1\not\succ t_2$, and $t_2\not\succ t_3$.
Consequently, $t_1\not\succ_0 t_3$, $t_1\not\succ t_3$, and thus
$t_1\not\succ' t_3$.\qed
\end{proof}

%We note that $\union{\succ}{\succ_{0}}$ (resp. $\pareto{\succ}{\succ_{0}}$)
%is trivially the preference relation
%which is closest to $\succ$ among those relations that contain $\union{\succ}{\succ_{0}}$
%(resp. $\pareto{\succ}{\succ_{0}}$).

Note that without the $0$-compatibility assumption, WOs are not closed with respect
to union and Pareto composition \cite{ChTODS03}.

For prioritized composition, we can relax the $0$-compatibility assumption.
This immediately follows from the fact that WOs are closed with respect
to prioritized composition \cite{ChTODS03}.
\begin{proposition}\label{prop:weak:weak:priority}
For every WO preference relations $\succ$ and $\succ_0$, the preference relation
$\prior{\succ_0}{\succ}$ is a WO.
\end{proposition}

A basic notion in utility theory is that of {\em representability} of preference relations
using numeric utility  functions:

\begin{definition}\label{def:represent}
A real-valued function $u$ over a schema $R$ 
{\em represents} a preference relation $\succ$ over $R$ iff
\[\forall t_1,t_2\ [t_1\succ t_2 \;{\rm iff}\; u(t_1)>u(t_2)].\]
Such a preference relation is called {\em utility-based}.
\end{definition}

Being a WO is a necessary condition for the existence of a numeric representation
for a preference relation. However, it is not sufficient for uncountable orders \cite{Fish70}.
It is natural to ask whether the existence of  numeric representations
for the preference relations $\succ$ and $\succ_0$ implies the existence of such 
a representation for the preference relation $\succ'=(\succ_0\theta\succ)$ where $\theta\in\{\cup,\rhd,\otimes\}$.
This is indeed the case. 

\begin{theorem}\label{th:utility}
Assume that $\succ$ and $\succ_{0}$ are WO preference
relations such that
\begin{enumerate}
\item $\succ$ and $\succ_{0}$ are $0$-compatible,
\item $\succ$  can be represented using a real-valued function $u$,
\item $\succ_{0}$ can be represented using a real-valued function $u_{0}$.
\end{enumerate}
Then $\succ'\, =\, \succ_0\theta\succ$, where $\theta\in\{\cup,\rhd,\otimes\}$,
is a WO preference relation that can be represented using any real-valued
function $u'$ such that for all $x$
\[u'(x)=a\cdot u(x)+b\cdot u_0(x)+c\]
where $a$ and $b$ are arbitrary positive real numbers.
\end{theorem}
\begin{proof}
Assume $x\succ' y$. Thus $x\succ_0 y$ or $x\succ y$.
If $x\succ_0 y$, then $u_0(x)>u_0(y)$.
Also, in this case $y\not\succ x$ because of $0$-compatibility.
This implies $u(x)\geq u(y)$.
Consequently, $u'(x) >u'(y)$.
The other case is symmetric.

Assume $u'(x) >u'(y)$. Thus $u_0(x)>u_0(y)$ or $u(x)> u(y)$.
In the first case, we get $x\succ_0 y$; in the second,
$x\succ y$. Consequently, $x\succ' y$.\qed
\end{proof}

Surprisingly, the $0$-compatibility requirement cannot in general be
replaced by $1$-compatibility if we replace $\cup$ by $\rhd$ in Theorem \ref{th:utility}.
\begin{example}
Consider the Euclidean space ${\cal R}\times{\cal R}$,
and the following orders:
\[\begin{array}{l}
(x,y)\succ_1 (x',y') \equiv x>x',\\
(x,y)\succ_2 (x',y') \equiv y>y',\end{array}\]
The orders $\succ_1$ and $\succ_2$ are $1$-compatible (but not $0$-compatible) WOs.
It is well known that their prioritized (also called {\em lexicographic})
composition is not representable using a utility function \cite{Fish70}.
\end{example}

Thus, preservation of {\em representability} is possible only 
under $0$-compatibility, in which case $\union{\succ_0}{\succ}=\prior{\succ_0}{\succ}=\pareto{\succ_0}{\succ}$
(Lemma \ref{l:coincide}).
(The results \cite{Fish70} indicate that for countable domains considered in this paper,
the prioritized composition of WOs, being a WO, is representable using a utility
function. However, that utility function is not definable 
in terms of the utility functions representing the given orders.)

We conclude this section by showing a general scenario in which the union of orders occurs in a natural way.
Assume that we have a numeric utility function $u$ representing a (WO) preference relation $\succ$.
The indifference relation $\sim$ generated by $\succ$ is defined as:
\[x\sim y\ \equiv\ u(x)=u(y).\]
Suppose that the user discovers that $\sim$ is too coarse and needs to be further refined.
This may occur, for example, when $x$ and $y$ are tuples and the function $u$ takes
into account only some of their components. Another  function $u_0$ may be defined
to take into account other components of $x$ and $y$ (such components are called
{\em hidden attributes} \cite{PuFaTo03}). The revising preference relation $\succ_0$
is now:
\[x\succ_0 y\ \equiv\ u(x)=u(y)\wedge u_0(x)>u_0(y).\]
It is easy to see that $\succ_0$ is an SPO $0$-compatible with $\succ$ (but not necessarily a WO). 
Therefore, by Theorem \ref{th:spo:union} the preference relation $\ \union{\succ}{\succ_0}\ $ is an SPO.
%(Actually, $\union{\succ}{\succ_0}$ is a WO. In general, it does not have a utility representation.)

\section{Incremental evaluation}\label{sec:incr}

\subsection{Query modification}

We show here how the already computed result of the original  preference query can be reused to make
the evaluation of the modified query  more efficient. We will use the following result.

\begin{proposition}{\rm \cite{ChTODS03}\label{prop:contain}}
If $\,\succ_{1}$ and $\succ_{2}$ are preference relations over a relation schema $R$ 
and \mbox{$\succ_{1}\, \subseteq\, \succ_{2}$}, 
then for all instances $r$ of $R$:
\begin{itemize} 
\item $\wnnw{\succ_2}(r)\,\subseteq\, \wnnw{\succ_1}(r);$
\item $\wnnw{\succ_2}(\wnnw{\succ_1}(r))\, =\, \wnnw{\succ_2}(r)$
if $\succ_1$ and $\succ_2$ are SPOs.
\end{itemize}
\end{proposition}

Consider the scenario in which we iteratively modify a given preference query by revising the preference relation
using only union in such a way that the revised preference relation is an SPO
(for example, if the assumptions of Theorem \ref{th:spo:union} are satisfied).
We obtain a sequence of preference relations 
$\succ_{1},\ldots,\succ_{n}$ such that $\succ_{1}\,\subseteq\cdots\subseteq\, \succ_{n}$.

In this scenario, the sequence of query results is:
\[r_0=r, r_1=\wnnw{\succ_1}(r),r_2=\wnnw{\succ_2}(r),\ldots,r_n=\wnnw{\succ_n}(r).\]

Proposition \ref{prop:contain} implies that the sequence $r_0,r_1,\ldots,r_n$ is decreasing:
\[r_0\supseteq r_1\supseteq\cdots\supseteq r_n\]
and that it can be computed incrementally:
\[r_1=\wnnw{\succ_1}(r_0),r_2=\wnnw{\succ_2}(r_1),\ldots,r_n=\wnnw{\succ_n}(r_{n-1}).\]
To compute $r_i$, there is no need to look at the tuples in \mbox{$r-r_{i-1}$,} nor to
recompute winnow from scratch. The sets of tuples $r_1,\ldots,r_n$ are likely to 
have  much smaller cardinality than $r_0=r$.

It is easy to see that the above comments apply to all cases where the revised
preference relation is a superset of the original preference relation.
Unfortunately, this is not the case for revisions that use  prioritized or Pareto composition.
However, given a specific pair of preference relations $\succ$ and $\succ_0$, one 
can still effectively check whether $\tc{\prior{\succ_0}{\succ}}$ or $\tc{\pareto{\succ_0}{\succ}}$
contains $\succ$ if the validity of preference formulas is decidable, as is the
case for ERO formulas (Proposition \ref{prop:NP}).

\ignore{
\begin{theorem}\label{th:incr:union}
For every preference relations $\succ$ and $\succ_0$ over $R$ which are SPOs
and every instance $r$ of $R$:
\[\wnnw{\tc{\union{\succ}{\succ_0}}}(r)=\wnnw{\tc{\union{\succ}{\succ_0}}}(\wnnw{\succ}(r)\cup\wnnw{\succ_0}(r))\]
and 
\[\wnnw{\tc{\prior{\succ_0}{\succ}}}(r)=\wnnw{\tc{\union{\succ}{\succ_0}}}(\wnnw{\succ_0}(r)).\]
So if $\wnnw{\succ}(r)$ and $\wnnw{\succ_0}(r)$ have already been computed, the modified query
$\wnnw{\tc{\union{\succ}{\succ_0}}}(R)$ can be evaluated over their union, instead of over the
entire relation instance $r$. Often, the results of winnow will be relatively small compared to the 
relation instance, and thus the above law could lead to sigificant savings.
}

\subsection{Database update}\label{sec:update}

In the previous section we studied query modification: the query is modified, while
the database remains unchanged. Here we reverse the situation: the query remains the same
and the database is updated. 

We consider first updates that are insertions of sets of tuples.
For a database relation $r$, we denote by $\add{r}$ the set of inserted tuples.
We show how the previous result of a given preference query can be reused to make
the evaluation of the same query in an updated database more efficient.

We first establish  the following result.

\begin{theorem}\label{th:delta}
For every preference relation $\succ$ over $R$ which is an SPO and every instance $r$ of $R$:
\[\wnnw{\succ}(r\cup\add{r})=\wnnw{\succ}(\wnnw{\succ}(r)\cup\add{r}).\]
\end{theorem}
\begin{proof}
Assume $t\not\in\wnnw{\succ}(\wnnw{\succ}(r)\cup\add{r})$.
Then either $t\not\in\wnnw{\succ}(r)\cup\add{r}$ or there exists $t'\in\wnnw{\succ}(r)\cup\add{r}$
such that $t'\succ t$.
In the first case, $t\not\in \wnnw{\succ}(r)$ and $t\not\in \add{r}$.
If $t\not \in r$ and $t\not\in\add{r}$, then $t\not\in \wnnw{\succ}(r\cup\add{r})$.
If there exists $t'\in \wnnw{\succ}(r)$ such that $t'\succ t$, then also
$t\not\in \wnnw{\succ}(r\cup\add{r})$.
In the second case, $t'\in\union{r}{\add{r}}$ and thus $t\not\in \wnnw{\succ}(r\cup\add{r})$.

Assume $t\not\in\wnnw{\succ}(r\cup\add{r})$.
Then either $t\not\in\union{r}{\add{r}}$ or there exists $t'\in\union{r}{\add{r}}$ such that
$t'\succ t$. 
In the first case, $t\not\in  \wnnw{\succ}(\wnnw{\succ}(r)\cup\add{r})$.
In the second case, if $t'\in \add{r}$, then $t\not\in  \wnnw{\succ}(\wnnw{\succ}(r)\cup\add{r})$.
So consider $t'\in r-\add{r}$.
If $t\in r$ but $t\not\in\add{r}$, then $t\not\in\wnnw{\succ}(r)\cup\add{r}$ and
$t\not\in  \wnnw{\succ}(\wnnw{\succ}(r)\cup\add{r})$.
If $t\in \add{r}$, then there exists $t''\in \wnnw{\succ}(r)$ such that $t''\succ t$.
($t''$ may be $t'$ or some element dominating $t'$.)
Therefore, in this case also $t\not\in  \wnnw{\succ}(\wnnw{\succ}(r)\cup\add{r})$.\qed
\end{proof}

Consider now the scenario in which we have a preference relation $\succ$, which is an SPO, and a sequence of relations 
\[r_0=r, r_1=r_0\cup\add{r_0},r_2=r_1\cup\add{r_1},\ldots,r_n=r_{n-1}\cup\add{r_{n-1}}.\]

Theorem \ref{th:delta} shows that
\[\begin{array}{l}
\wnnw{\succ}(r_1)=\wnnw{\succ}(\wnnw{\succ}(r_0)\cup\add{r_0})\\
\wnnw{\succ}(r_2)=\wnnw{\succ}(\wnnw{\succ}(r_1)\cup\add{r_1})\\
\ldots\\
\wnnw{\succ}(r_{n})=\wnnw{\succ}(\wnnw{\succ}(r_{n-1})\cup\add{r_{n-1}}).
\end{array}\]

Therefore, each subsequent evaluation of winnow can reuse the result of the previous one.
This is advantageous because winnow returns a subset of the given relation and this subset
is often much smaller than the relation itself.

Clearly, the algebraic law, stated in Theorem \ref{th:delta}, can be used together with other, well-known laws
of relational algebra and the laws specific to preference queries \cite{ChTODS03,KiHa03}
 to produce a variety of rewritings of a given preference query.
To see how a more complex preference query can be handled, let's consider
the query consisting of winnow and selection, $\wnnw{\succ}(\sigma_{\alpha}(R))$.
We have
\[\wnnw{\succ}(\sigma_{\alpha}(r\cup\add{r}))=\wnnw{\succ}(\sigma_{\alpha}(r)\cup\sigma_{\alpha}(\add{r}))=
\wnnw{\succ}(\wnnw{\succ}(\sigma_{\alpha}(r))\cup\sigma_{\alpha}(\add{r}))\]
for every instance $r$ of $R$.
Here again, one can use the previous result of the query, $\wnnw{\succ}(\sigma_{\alpha}(r))$, to make
its current evaluation more efficient.
Other operators that distribute through union, for example projection and join, can be handled in the same way.

Next, we consider updates that are deletions of sets of tuples.
For a database relation $r$, we denote by $\del{r}$ the set of deleted tuples.

\begin{theorem}\label{th:delete}
For every preference relation $\succ$ over $R$ and every instance $r$ of $R$:
\[\wnnw{\succ}(r)-\del{r}\subseteq \wnnw{\succ}(r-\del{r}).\]
\end{theorem}

Theorem \ref{th:delete} gives  an incremental way to compute an approximation of winnow from below.
It seems that in the case of deletion there cannot be an exact law along the lines of Theorem \ref{th:delta}.
This is because the deletion of some tuples from the original database may promote some originally
dominated (and discarded) tuples into the result of winnow over the updated database.

\begin{example}
Consider the following preference relation $\succ=\{(a,b_1),\ldots,(a,b_n)\}$
and the database $r=\{a,b_1,\ldots,b_n\}$. Then 
$\wnnw{\succ}(r)=\{a\}$ but 
\[\wnnw{\succ}(r-\{a\})=\{b_1,\ldots,b_n\}.\]
\end{example}

\section{Finite restrictions of preference relations}\label{sec:finite}

\subsection{Restriction}
It is natural to consider {\em restrictions} of preference relations to given database instances \cite{ToCi02}.
\begin{definition}\label{def:finite}
Let $r$ be an instance of a relation schema $R$ and $\succ$ a preference relation over $R$.
The {\em restriction $\relpref{\succ}{r}$ of $\succ$ to $r$} is a preference relation over $R$, defined as
\[\relpref{\succ}{r}\,=\ \succ\cap\ r\times r.\]
\end{definition}

We write $(x,y)\in \relpref{\succ}{r}$ instead of $x\relpref{\succ}{r}y$ for greater readability.

The advantage of using $\relpref{\succ}{r}$ instead of $\,\succ$ comes from the fact that
the former depends on the database contents and can have stronger properties than the latter.
For example, $\relpref{\succ}{r}$ may be an SPO, while $\succ$ is not.
Similarly, $\relpref{\succ}{r}$ may be $i$-compatible
with $\relpref{\succ_0}{r}$, while $\succ$ is not $i$-compatible with $\succ_0$.
Therefore, restrictions could be used instead of preference relations in the revision process.
%On the other hand, $\succ$ makes more elaborate use of the preference information than $\relpref{\succ}{r}$
%and does not require adaptation if the input database changes.

The following is a basic property of  restriction. It says that the restriction to an instance does
not affect the result of winnow {\em over the same instance}, so the restriction can be used
in place of the original preference relation.
\begin{theorem}\label{th:finite}
Let $r$ be an instance of a relation schema $R$ and $\succ$ a preference relation over $R$.
Then
\[\wnnw{\relpref{\succ}{r}}(r)=\wnnw{\succ}(r).\]
\end{theorem}
\begin{proof}
We have $\relpref{\succ}{r}\subseteq r$ and thus $\wnnw{\succ}(r)\subseteq \wnnw{\relpref{\succ}{r}}(r)$.
In the other direction, assume $t\not\in\wnnw{\succ}(r)$.
If $t\not\in r$, $t\not\in\wnnw{\relpref{\succ}{r}}(r)$.
If $t\in r$ and there exists $t'\in r$ such that $t'\succ t$, then also $(t',t)\in \relpref{\succ}{r}$
and $t\not\in\wnnw{\relpref{\succ}{r}}(r)$.\qed
\end{proof}

We also establish that restriction distributes over the preference composition operators.

\begin{theorem}\label{th:distr}
If $r$ is an instance of a relation schema $R$, $\theta\in\{\cup,\rhd,\otimes\}$, and 
$\succ$ and $\succ_0$ are preference relations over $R$, then
\[\relpref{\succ_0\theta \succ}{r}=\relpref{\succ_0}{r}\theta\relpref{\succ}{r}.\]
\end{theorem}
\begin{proof}
We prove this result for $\theta=\rhd$. The other cases are similar.

We have the following equivalences:
\[\begin{array}{l}
(x,y)\in \relpref{\succ_0}{r}\rhd\relpref{\succ}{r}\equiv\\
(x,y)\in\relpref{\succ_0}{r}\vee (y,x)\not\in \relpref{\succ_0}{r}
\wedge (x,y)\in\relpref{\succ}{r}\equiv\\
x\succ_0 y \wedge x\in r\wedge y\in r\vee (y\not\succ_0 x \vee x\not\in r 
\vee y \not\in r)\wedge x\succ y \wedge x\in r \wedge y\in r\equiv\\
x\succ_0 y \wedge x\in r\wedge y\in r\vee y\not\succ_0 x 
\wedge x\succ y \wedge x\in r \wedge y\in r\equiv\\
(x\succ_0 y \vee y\not\succ_0 x \wedge x\succ y)\wedge x\in r \wedge y\in r\equiv\\
(x,y)\in \relpref{\succ_0\rhd \succ}{r}.
\end{array}\]
\end{proof}

The preference revision studied earlier in this paper typically involved the computation of the
of the revised preference relation defined as 
the transitive closure $\tc{\succ_0\theta \succ}$, where $\theta\in\{\cup,\rhd,\otimes\}$,
$\succ$ is the original preference relation, and $\succ_0$ is the revising preference relation.
We study several different ways of imposing the restriction of preferences to a relation instance.
We consider the following preference relations:
\[\begin{array}{lcl}
\succ_1&=&\tc{\succ_0\theta \succ},\\
\succ_2&=&\relpref{\tc{\succ_0\theta \succ}}{r},\\
\succ_3&=&\tc{\relpref{\succ_0\theta \succ}{r}},\\
\succ_4&=&\tc{\relpref{\succ_0}{r}\theta\relpref{\succ}{r}}.\\
\end{array}\]

We establish now some fundamental relationships between the preference relations
$\succ_1, \succ_2, \succ_3$, and $\succ_4$.

\begin{theorem}\label{th:contain}
Let $\theta\in\{\cup,\rhd,\otimes\}$, and $\succ$ and $\succ_0$ be preference relations over a schema 
$R$. Then for every instance $r$ of $R$: 
\[\succ_4\ =\ \succ_3\ \subseteq\ \succ_2\ \subseteq\ \succ_1,\]
and there are relation instances for which the containments are strict.
\end{theorem}
\begin{proof}
The equality of $\succ_4$ and $\succ_3$ follows from Theorem \ref{th:distr}.
For $\succ_3\ \subseteq\ \succ_2$,
we have that 
\[\relpref{\succ_0\theta \succ}{r}\ \subseteq\ \succ_0\theta \succ,\]
and
\[\relpref{\succ_0\theta \succ}{r}\subseteq r\times r.\]
Thus 
\[\succ_3=\tc{\relpref{\succ_0\theta \succ}{r}}\subseteq r\times r,\]
and 
\[\succ_3\subseteq \tc{\succ_0\theta \succ}\cap r\times r=\succ_2.\]
The containment $\succ_2\ \subseteq\ \succ_1$ follows from the definition of the
restriction.

An example where $\succ_3\ \subset\ \succ_2\ \subset\ \succ_1$ is as follows.
Let $\succ=\{(a,b)\}$, $\succ_0=\{(b,c)\}$, $r=\{a,c\}$.
Then $\relpref{\succ}{r}=\relpref{\succ_0}{r}=\emptyset$.
Thus also $\relpref{\succ_0\theta \succ}{r}=\relpref{\succ_0}{r}\ \theta\ \relpref{\succ}{r}=\emptyset$,
and $\succ_3=\emptyset.$
On the other hand, $\succ_1=\{(a,b),(b,c),(a,c)\}$
and $\succ_2=\{(a,c)\}$.\qed
\end{proof}

\begin{corollary}\label{cor:finite}
Let $\theta\in\{\cup,\rhd,\otimes\}$, and $\succ$ and $\succ_0$ be preference relations over a schema 
$R$. Then for every instance $r$ of a $R$:
\[\wnnw{\succ_1}(r)=\wnnw{\succ_2}(r)\subseteq \wnnw{\succ_3}(r)=\wnnw{\succ_4}(r),\]
and for some cases the containment is strict.
\end{corollary}
\begin{proof}
Follows from Theorem \ref{th:finite} and Theorem \ref{th:contain}.
In the example given in the proof of Theorem \ref{th:contain},
we obtain $\wnnw{\succ_2}(r)=\{a\}$ and $\wnnw{\succ_3}(r)=\{a,c\}$.\qed
\end{proof}

We study now the order-theoretic properties of restriction.

\begin{theorem}\label{th:restr:spo}
Let $\theta\in\{\cup,\rhd,\otimes\}$, and $\succ$ and $\succ_0$ be preference relations over a schema 
$R$. Then for every instance $r$ of $R$,
$\succ_1$ is an SPO implies that $\succ_2$ is an SPO, which implies that $\succ_3$ is an SPO.
There are cases in which the reverse implication does not hold.
\end{theorem}
\begin{proof}
Because $\succ_2\ \subseteq\ \succ_1$, $\succ_2$ is irreflexive.
Assume that $x\succ_2 y$ and $y\succ_2 z$.
Then $x\succ_1 y$, $y\succ_1 z$, $x\in r$, $y\in r$, and $z\in r$.
Therefore, $x\succ_1 z\wedge x\in r\wedge z\in r$, and $x\succ_2 z$.

The preference relation $\succ_1=\{(a,a)\}$ is not an SPO (and can be obtained
from some preference relations $\succ_0$ and $\succ$ using any composition operator). However, 
its restriction $\succ_2=\relpref{\succ_1}{r}$ for $r=\{b\}$ is empty,
and thus an SPO.

Assume now $\succ_0=\{(a,b)\}$ and $\succ=\{(b,a)\}$.
Consider $\theta=\cup$ and $r=\{b\}$. Thus,
$\succ_1=\{(a,b),(b,a),(a,a),(b,b)\}$ and $\succ_2=\{(b,b)\}$
(so it is not an SPO).
On the other hand, $\relpref{\succ_0\cup\succ}{r}=\emptyset$
and $\succ_3=\emptyset$, too.
Similar examples can be constructed for the other composition operators.\qed
\end{proof}

Unfortunately, for weak orders there is no property analogous to Theorem \ref{th:restr:spo}.

Subsequently, we examine the impact of restriction on compatibility.

\begin{theorem}
Let $\succ$ and $\succ_0$ be preference relations over a schema 
$R$. Then for every instance $r$ of a relation schema $R$ and every $i=0,1,2$
if $\succ$ is $i$-compatible with $\succ_0$, then
$\relpref{\succ}{r}$ is $i$-compatible with $\relpref{\succ_0}{r}$.
There are cases in which the reverse implications do not hold.
\end{theorem}
\begin{proof}
For $0$-compatibility the situation is clear.
If there are no $0$-conflicts between $\succ$ and $\succ_0$,
then there are no $0$-conflicts between 
$\relpref{\succ}{r}$ and $\relpref{\succ_0}{r}$.
However, for higher-level conflicts, the situation is more
complicated.

Assume now that $\succ$ is $1$-compatible with $\succ_0$ and 
consider a $1$-conflict between $\relpref{\succ}{r}$ and 
$\relpref{\succ_0}{r}$. Then there are elements $t_1,t_2,s_1,\ldots,s_k$
of $r$ such that
\[(t_1,t_2)\in \relpref{\succ_0}{r}, (t_2,s_1)\in \relpref{\succ}{r}, \ldots,
(s_k,t_1)\in\relpref{\succ}{r},\]
and 
\[(t_1,s_k)\not\in\relpref{\succ_0}{r},\ldots, (s_1,t_2)\not\in\relpref{\succ_0}{r}.\]
Consider now any two elements $x$ and $y$ among $t_1,t_2,s_1,\ldots,s_k$
such that $(x,y)\in\relpref{\succ}{r}$ (resp.$(x,y)\in \relpref{\succ_0}{r}$).
Clearly then also $x\succ y$ (resp., $x\succ_0 y$).
Assume  $(x,y)\in\relpref{\succ}{r}$ and $(y,x)\not\in\relpref{\succ_0}{r}$.
Thus $y\not\succ_0 x$. So we obtain a $1$-conflict between the
preference relations $\succ$ and $\succ_0$.
$2$-conflicts are analyzed in the same fashion.

To see that the lack of $1$-conflicts between
$\relpref{\succ}{r}$ and $\relpref{\succ_0}{r}$ does not
imply the lack of $1$-conflicts between $\succ$ and $\succ_0$,
consider 
\[\succ_0=\{(c,a)\}\]
\[ \succ=\{(a,b),(b,c),(a,c)\},\]
and $r=\{(a,c)\}$.
Then $\relpref{\succ}{r}=\{(a,c)\}$ and $\relpref{\succ_0}{r}=\{c,a\}$.
There are no $1$-conflicts between
$\relpref{\succ}{r}$ and $\relpref{\succ_0}{r}$
but there is a $1$-conflict between $\succ$ and $\succ_0$.
Analogous examples can be constructed for other kinds of conflicts.\qed
\end{proof}

Finally, we  compare the computational properties of
$\succ_1,\succ_2$ and $\succ_3$.
The preference relation $\succ_1$ is recomputed only after preference
revisions.
The relation $\succ_2$ is recomputed after every revision and every
database update.
The recomputation after an update uses $\succ_1$ as a selection condition
applied to $r\times r$ (where $r$ is the current relation instance).
The relation $\succ_3$ is also recomputed after every revision and every
database update.
However, in the latter case the computation is more involved than for
$\succ_2$, because transitive closure of a finite binary relation needs to be computed.
Overall, $\succ_1$ represents the most stable and comprehensive preference information.
Even if $\succ_2$ is stored, $\succ_1$ needs to be kept up-to-date
after preference revisions, since it is used in the recomputation of
$\succ_2$ after an update.
The preference relation $\succ_3$ can be stored, revised, and updated without any
reference to $\succ_1$. However, in this case some preference information is lost,
c.f., Corollary \ref{cor:finite}.

\ignore{
Thus $\wnnw{\succ}(r)=\wnnw{\relpref{\succ}{r}}(r)=\{a,c\}$.
Consider revision using union, as in Theorem \ref{th:spo:union}.
The revised preference relation $\succ_1=\tc{\union{\succ}{\succ_0}}=\{(a,b),(b,c),(a,c)\}$.
On the other hand, $\relpref{\succ}{r}=\relpref{\succ_0}{r}=\emptyset$.
Thus the revised preference relation $\succ_2=\tc{\union{\relpref{\succ}{r}}{\relpref{\succ_0}{r}}}=\emptyset$.
After the revision, $\wnnw{\succ_1}(r)=\{a\}$ and $\wnnw{\succ_2}=\{a,c\}$.
So in the latter  case revision has no impact on preference.
We also note that $\relpref{\tc{\union{\succ}{\succ_0}}}{r}\not=\tc{\union{\relpref{\succ}{r}}{\relpref{\succ_0}{r}}}$,
and thus the correspondence between the unrestricted and the restricted preference relations no longer holds after
the revision.
}

\subsection{Non-intrinsic preferences}
Non-intrinsic preference relations are
defined using formulas that refer not only to built-in predicates. 
\begin{example}\label{ex:stored}
The following preference relation is not intrinsic:
\[x\succ_{\mathit Pref} y\equiv (x,y)\in {\mathit Pref}\]
where ${\mathit Pref}$ is a database relation.
One can think of such a relation as representing {\em stored} preferences.
\end{example}

Revising non-intrinsic preference relations looks problematic.
First, it is typically not possible to establish the simplest order-theoretic
properties of such relations. 
For instance, in Example \ref{ex:stored} it is not possible to determine  the
irreflexivity or transitivity of $\succ_{\mathit Pref}$ on the basis of its definition.
Whether such properties are satisfied depends on the contents of the database
relation ${\mathit Pref}$.
Second, the transitive closure of a non-intrinsic preference relation may fail to be 
expressed as a finite formula. Again, Example \ref{ex:stored} can be used to illustrate this point.

However, it seems that restriction may be able to alleviate the above problems.
Suppose $\succ$ is the original and $\succ_0$  the revising preference relations.
Computing $\tc{\succ_0\cup\succ}$  may be infeasible,
as indicated above. But computing 
$\tc{\relpref{\succ_0\cup \succ}{r}}$ is not difficult,
as $\relpref{\succ_0\cup \succ}{r}$ is computed by the first-order query
\[(x\succ_0 y \vee x \succ y) \wedge x\in R \wedge y\in R.\]
For other composition operators, the same approach also works because they are, like
union, defined in a first-order way.

\section{Weak-order extensions}\label{sec:ext}

Theorems \ref{th:spo:weak:priority} and \ref{th:weak:weak}, and
Proposition \ref{prop:weak:weak:priority} demonstrate that for weak
orders one can prove stronger properties about revisions than for
general partial orders. The $0$-compatibility or the interval order requirements may be
relaxed, and the transitive closure computation may no longer be
necessary.

So it would be advantageous to work with weak orders. Such orders can, for example, be obtained
as {\em extensions} of the given SPOs. 
We show here how to express the construction of weak order extensions using Datalog$\neg$ rules
\cite{AbHuVi95} and the Rule Algebra \cite{ImNa88}.
Although not much can be shown in that framework about WO extensions of arbitrary SPOs,
the construction of WO extensions of interval orders (IOs) can be guaranteed to terminate.

\subsection{Rules}
We define the {\em application} $r(X)$ of a rule $r$ to an input  set of facts $X$ in the standard way.
\begin{definition}\label{def:rule}
Assume $r$ is of the form
\[A\leftarrow B_1,\ldots,B_n,\neg C_1,\ldots,\neg C_m.\]
Then $r(X)$ consists of all the facts $\tau(A)$ such that $\tau(B_i)\in X$,
$i=1,\ldots,n$, and $\tau(C_j)\not\in X$, $j=1,\ldots,m$,
where $\tau$ is a ground substitution.
In an {\em inflationary} application $r(X)$ is added to $X$.
%A rule application {\em saturates} for $X$ if $r(X)\subseteq X$.
\end{definition}

In this paper, we are dealing with infinite sets of facts represented by constraints.
However, the above definition of rule application still applies.
From this definition, we can obtain a more operational definition that will tell us
how to construct the constraints in the head of the rule $r$ from the constraints in the body
\cite{CDB00}.

Assume that each goal $B_i$, $i=1,\ldots,n$ is described by a constraint $\beta_i$
and each goal $C_j$, $j=1,\ldots,m$ by a constraint $\gamma_j$.
Also denote by $V$ the set of variables that occur only in the body of $r$.
Then $A$ is described by the formula 
\[\exists\, V.\ \beta_1\wedge\cdots\wedge \beta_n\wedge \neg\gamma_1\wedge\cdots\wedge\neg\gamma_m.\]
from which negation and quantifiers have been eliminated.

\cite{ImNa88} present a language called Rule Algebra (RA) which allows rule composition.
The syntax of RA expressions is defined as follows:
\[Expr ::= r\, |\, Expr\,;\,Expr\, |\, Expr\, \cup\, Expr\, |\, Expr^+,\]
where $r$ is a single rule.
The symbol ``$;$'' denotes sequential and ``$\cup$'', parallel composition.
The superscript ``$+$'' denotes unbounded iteration.

The application  of RA expressions is defined as follows \cite{ImNa88}:
\begin{itemize}
\item for a single rule it is defined as in Definition \ref{def:rule},
\item $(F_1;F_2)(X)\doteq F_2(F_1(X))$,
\item $(F_1\cup F_2)(X)\doteq F_1(X)\cup F_2(X)$,
\item $F^+(X)\doteq\bigcup_{i>0}F^i(X)$.
\end{itemize}

%The saturation for sequential and parallel composition is defined in the same way
%as for single rules (Definition \ref{def:rule}).
%For iteration, $F^+(X)$ saturates for if for some $i$, F^i(X)

%A Rule Algebra expression saturates if $E(X)\subseteq X$.
Like rule application, the application of RA expressions  comes in two different variants:
inflationary and non-inflationary.

Rule Algebra can be implemented directly. However,
\cite{ImNa88} show also how to map Rule Algebra expressions to a class of {\em locally-stratified}
logic programs \cite{Prz88}. This class requires a limited use of function symbols to implement
counters. 

\subsection{Strict partial orders}

\cite{Fish85} presents a construction of a WO extension  of a finite  SPO.
It is based on a very simple intuition.

Assume we are given that $x\succ y$ and $y\sim z$, or $x\sim y$ and $y\succ z$.
In a weak order one needs to be able to have also $x\succ z$ in both cases (see Proposition \ref{prop:weak}).
Therefore, one could produce a WO extension $\succ'$ of a given SPO $\succ$ by supporting
the {\em derivation} of the implied order relationships. Clearly, such derivation should
avoid contradiction ($x\succ' y$ and $y\succ' x$).
\begin{example}\label{ex:ext:1}
Consider the following order $\succ=\{(a,c),(b,d)\}$.
Thus $a\sim d$ and $b\sim c$. So w could derive $a\succ' b$ and $b\succ' a$,
a contradiction.
\end{example}

We construct an extension $\succ'$ of a given SPO $\succ$
using a set of rules. Unfortunately, for infinite orders the construction
does not always produce a weak order.
The input preference relation $\succ$ is described using a set of facts of the relation $T$ of 
arity $2n$ where $n$ is the arity of the database relation over which $\succ$ is defined.
The output preference relation $\succ'$ is also described as a set of facts of the relation $T$ but those
facts are computed using rule application.
%The symbol ``$\neg$'' stands for negation.

First, we have two rules $P_{11}$ and $P_{12}$ for deriving new order relationships:
\[P_{11}:\ T(x,z) \leftarrow T(x,y) \wedge \neg T(z,y)\wedge \neg T(y,z).\]
\[P_{12}:\ T(x,z) \leftarrow T(y,z)\wedge \neg T(x,y)\wedge \neg T(y,x).\]
Second, we have the conflict removal rule $P_2$:
\[P_2:\ T(x,y) \leftarrow T(x,y)\wedge\neg T(y,x).\]

We note that
the rules $P_{11},P_{12},P_2$ need to be applied in a specific order.
We use the following  Rule Algebra expression $E_1$ \cite{ImNa88,AbHuVi95}
\[E_1=((P_{11}\ \cup\ P_{12})\ ;\ P_2)^+,\]
applied to the input preference relation.
In the rule $P_2$ and the expression $E_1$, the desired semantics is
non-inflationary because we want to eliminate conflicts.

\ignore{
It does not seem possible to map the above expression $E$ 
to a program in one of the variants of (function-free) Constraint Datalog$\neg$ in such a way that the desired
semantics is preserved. (For such variants, termination of programs has been studied
\cite{KaKuRe95,TomCDB97}.) In particular, the rule $P_2$ has non-inflationary semantics.
Also, the well-founded model approach \cite{VGRoSc91} does not seem appropriate.
}

\begin{example}
Consider the preference relation $\succ=\{(a,c),(b,d)\}$ from Example \ref{ex:ext:1}.
Applying the rules $P_{11}$ and $P_{12}$ we obtain the relation
\[T(x,y)\equiv x=a\wedge y\not= a\vee x=b\wedge y\not= b\vee x\not=c \wedge y=c\vee x\not=d\wedge y=d.\]
This is not an SPO because, for example,  we have $T(a,b)$ and $T(b,a)$.
Applying the rule $P_2$, the conflict is removed, yielding
\[\begin{array}{lcl}
T(x,y)&\equiv &x=a\wedge y\not= a\wedge y\not= b\vee x=b\wedge y\not= b\wedge y\not= a\\
&&\vee x\not=c\wedge x\not=d\wedge y=c\vee x\not=c\wedge x\not=d\wedge y=d.
\end{array}\]
which is a weak order.
Thus, no further iterations are necessary.
\end{example}

Denote by $T_i$ the preference relation obtained at the end of the $i$-th stage
in the computation of $E_1$.
Clearly, if $T_i$ is a weak order, then nothing new is produced at the next stage, i.e., $T_{i+1}=T_i$.
However, the reverse implication does not have to hold for arbitrary SPOs.
Therefore, in each stage $i$, $T_i$ needs to be separately checked for the weak order
property (Proposition \ref{prop:NP} implies that the appropriate properties are decidable
under the assumption that the input preference relation is described by an  ERO preference formula).
%In some cases, the number of iterations can be bounded in advance.
%For example, if $\succ$ is a semi-order, one iteration suffices \cite{Luc56}.
\begin{example}\label{ex:hole}
Consider the following rational-order preference relation $\succ$ adapted from \cite{Fish85}:
\[x\succ y \equiv x>y\wedge x\not=0 \wedge y\not=0.\]
The corresponding indifference relation $\sim$ is defined as
\[x\sim y\equiv x=y \vee x=0 \vee y=0.\]
The relation $\succ$ is not a weak order but even the first iteration of the above rules fails to
produce anything new.
Consider any rational number $b\not=0$. There are numbers $a$ and $c$ such that  $a>b$, $b>c$, $a\sim 0$ and $c\sim 0$.
So on the one hand we have initially $T(b,c)$, $\neg T(c,0)$ and $\neg T(0,c)$, and
applying the rule $P_{11}$ we get $T(b,0)$.
But on the other hand we have $T(a,b)$, $\neg T(a,0)$ and $\neg T(0,a)$.
Applying the rule $P_{12}$ we get $T(0,b)$.
Therefore, the rule $P_2$ does not derive $T(b,0)$, $T(0,b)$, or any other new fact.
\end{example}

It is an open question what kind of properties a preference relation
should satisfy so that the condition $T_{i+1}=T_i$ implies the weak order
property. \cite{Fish85} shows that such an implication holds for SPOs
over finite domains. Therefore, it also holds for finite restrictions of arbitrary SPOs
(Section \ref{sec:finite}).
For a finite restriction $\relpref{\succ}{r}$ a different way for constructing a weak
order extension of $\relpref{\succ}{r}$ is available through the use of {\em ranking} \cite{ChTODS03}.
The ``best'' tuples -- those in $\wnnw{\succ}(r)$ --  receive rank 1, the ``second-best''
rank 2 etc. Then the weak order extension $\succ'$ is defined as
\[x\succ' y \equiv rank(x)<rank(y).\]

%Another direction is to develop a different construction of weak-order extensions.

\subsection{Interval orders}

For interval orders, we can show stronger results about constructing WO extensions.
We still use the Datalog$\neg$/Rule Algebra  framework but instead of the expression $E_1$
we use the following expression $E_2$:
\[E_2=\ (P_{11}\ ;\ P_{12})^+.\]

We will see that for $E_2$ the inflationary and non-inflationary semantics coincide.

For simplicity, we identify here a preference relation with the set of facts of the $T$ predicate describing it.

\begin{example}\label{ex:hole:1}
Consider Example \ref{ex:hole}.
Applying the rule $P_{11}$ to the preference relation $\succ$ from this example (which is an interval order)
yields the following preference relation $\succ'$:
\[x\succ' y \equiv x>y \wedge x\not=0\wedge y\not=0 \vee x\not=0\wedge y=0.\]
This relation is a total order, and thus also a weak order.
\end{example}

\begin{lemma}\label{lem:inflationary}
For every irreflexive preference relation $X$, $X\subseteq P_{11}(X)$,
$X\subseteq P_{12}(X)$, and $X\subseteq P_{12}(P_{11}(X))$.
\end{lemma}

\begin{lemma}\label{lem:preserve}
Assume $X$ is an interval order preference relation.
Then $P_{11}(X)$ and $P_{12}(X)$ are interval order preference relations.
\end{lemma}
\begin{proof}
WLOG, consider $Y=P_{11}(X)$. Clearly, $Y$ is irreflexive.
For transitivity, consider $T(x,y)\in Y$ and $T(y,z)\in Y$.
Then there is a $z'$ such that $T(x,z')\in X$, $T(z',y)\not\in X$,
and $T(y,z')\not\in X$.
Similarly, there is a $z''$ such that $T(y,z'')\in X$, $T(z'',z)\not\in X$,
and $T(z,z'')\not\in X$.
Because $X$ is an interval order, we have $T(x,z'')\in X$ or $T(y,z')\in X$.
Assume the former. Then $T(x,z)\in Y$. The preservation of the interval order
condition can be shown in a similar way.\qed
\end{proof}

\ignore{
To see that $Y$ is an interval order, consider $T(x,y)\in Y$ and $T(z,w)\in Y$.
Then there is a $z'$ such that $T(x,z')\in X$, $T(z',y)\not\in X$,
and $T(y,z')\not\in X$.
Similarly, there is a $z''$ such that $T(z,z'')\in X$, $T(z'',w)\not\in X$,
and $T(w,z'')\not\in X$.
Thus, $T(x,z'')\in X$ or $T(z,z')\in X$.}

\begin{lemma}\label{lem:sat}
%The expression $E_2$ saturates for a given interval order preference relation $X$ iff $X$ is a weak order.
Let $F=(P_{11};P_{12})$ and $Y$ be an SPO. Then $F(Y)\subseteq Y$ iff $Y$ is a WO.
\end{lemma}
\begin{proof}
If $Y$ is a WO, then 
\[Y=P_{11}(Y)=P_{12}(P_{11}(Y)).\]
If $Y$ is not a WO but an SPO, then there are $x$, $y$ and $z$ such that
$T(x,y)\in Y$, $T(x,z)\not\in Y$, $T(z,x)\not\in Y$, $T(y,z)\not\in Y$ and $T(z,y)\not\in Y$.
Thus $T(x,z)\in P_{11}(Y)$ and by Lemma \ref{lem:inflationary}, $T(x,z)\in P_{12}(P_{11}(Y))$.
Thus $P_{12}(P_{11}(Y))\not\subseteq Y$.\qed
\end{proof}

The following theorem shows that finite termination of the evaluation of $E_2$ 
is equivalent to the weak order property.

\begin{theorem}\label{th:interval:sat}
Let $X$ be an IO.
For every $i>0$, 
$E_2(X)=(P_{11};P_{12})^+(X)$ equals $(P_{11};P_{12})^i(X)$ iff $(P_{11};P_{12})^i(X)$
is a WO.
\end{theorem}
\begin{proof}
Follows from Lemmas \ref{lem:inflationary}, \ref{lem:preserve}, and  \ref{lem:sat}.
Note that for $j<i$, $(P_{11};P_{12})^j(X)\subseteq(P_{11};P_{12})^i(X)$.
It is essential that the given preference relation be an IO. Otherwise,
an application of $P_{11};P_{12}$ may produce preference relations which are not SPOs
and the equivalence in Lemma \ref{lem:sat} may stop to hold.
\qed
\end{proof}

To explore the possible implementations of the Rule Algebra expression $E_2$,
we note first that Lemma \ref{lem:inflationary} implies that for the rules
$P_{11}$ and $P_{12}$ inflationary and non-inflationary semantics coincide. Therefore, we can use 
inflationary or non-inflationary languages for the implementation of $E_2$.
\cite{AbHuVi95} indicate that Rule Algebra expressions can be translated to
Inflationary Datalog$\neg$  \cite{GuSh86}, a variant of
Datalog that allows unstratified negation (necessary here because of
the rules $P_{11}$ and  $P_{12}$) at the price of having
inflationary semantics.  
It is clear that Inflationary Datalog$\neg$
programs terminate on finite inputs.  However, preference relations
are typically infinite. Still, they are finitely representable using
preference formulas, and thus we are dealing with the problem of
termination of Inflationary Constraint Datalog$\neg$ programs.
Fortunately, there are positive results established in this area in
\cite{KaKuRe95}, which, together with Theorem \ref{th:interval:sat}, imply the following:
\begin{theorem}
Every interval order preference relation $\succ$, defined using an ERO
formula, has a weak order extension $\succ'$, defined using an ERO formula.
The formula defining $\succ'$  can be computed in exponential
time.
\end{theorem}

\section{Related work}\label{sec:related}

\subsection{Preference change}
\cite{Han95} presents a general framework for modeling change in preferences.
Preferences are represented syntactically using sets of ground preference formulas,
and their semantics is captured using sets of preference relations.
Thanks to the syntactic representation preference revision is treated similarly, though not
identically, to belief revision \cite{GaRo95}, and some axiomatic properties of preference revisions are identified.
The result of a revision is supposed to be 
minimally different from the original preference relation (using a notion of minimality based on symmetric difference)
and satisfy some additional background postulates, for example specific
order axioms. \cite{Han95} does not address the issue of constructing or defining revised relations, 
nor 
does it study the properties of specific classes of preference relations.
On the other hand, \cite{Han95} discusses also preference contraction, and domain expansion and shrinking.

In our opinion, there are several fundamental differences between belief and preference revision.
In belief revision, propositional theories are revised with propositional formulas, yielding
new theories. In preference revision, binary preference relations are revised with other
preference relations, yielding new preference relations.
Preference relations are single, finitely representable (though possibly infinite) first-order structures, 
satisfying order axioms.
Belief revision focuses on axiomatic properties of belief revision operators
and various notions of revision minimality. Preference revision focuses
on axiomatic, order-theoretic properties of revised preference relations  and the definability of such relations
(though still taking revision minimality into account).

\cite{Wil97} considers revising a ranking (a WO) of a finite set of tuples with new
information, and shows that a  new ranking, satisfying the AGM belief revision postulates \cite{GaRo95}, can be 
computed in a simple way. 
\cite{RevICDT97} describes a number of different revision operators for constraint databases.
However, the emphasis is on the axiomatic properties of the operators, not on the definability
of revised databases.
\cite{PuFaTo03} formulates various scenarios of preference revision and does not contain any formal framework.
\cite{Wong94} studies revision and contraction of finite WO preference relations by single pairs $t_1\succ_0 t_2$.
\cite{Freu04} describes minimal change revision of {\em rational} preference relations between propositional formulas.
%We are not aware of any work on revising infinite preference relations.

\ignore{
To apply the results of the present  paper to CP-nets, the following steps need to be done:
\begin{enumerate}
\item compute the Boolean constraint representation of the preference relation 
$TC(\succ_{C_M})$ corresponding to the revised  CP-net $M$ and the preference relation
$TC(\succ_{C_{M_0}})$ corresponding to the revising CP_net $M_0$;
\item determine whether any result of the present paper is applicable and if this is the case,
construct an appropriate least refinement;
\item map the obtained preference relation to a new CP-net.
\end{enumerate}
It remains to be seen whether the last step above can always be done and
how efficient the Constraint Datalog  computation could be made in this context.
}

\subsection{Preference queries}
Two different approaches to preference queries have been pursued
in the literature: qualitative and quantitative. In the {\em
qualitative} approach, 
preferences are
specified  using binary {\em preference
relations}  \cite{LaLa87,GoJaMa01,ChEDBT02,ChTODS03,Kie02,KiKo02}. 
In the  {\em quantitative} utility-based approach, preferences are represented using
{\em numeric utility functions} \cite{AgWi00,HrPa04}, as shown in Section \ref{sec:preserve}.
The qualitative approach is strictly more general than the
quantitative one, since one can define preference relations in
terms of utility functions. However, only WO
preference relations can be represented by numeric utility
functions \cite{Fish70}. 
Preferences that are not WOs are common in database applications,
c.f., Example \ref{ex:car}.
\begin{example}
There is no utility
function that captures the preference relation described in Example \ref{ex:car}.
Since there is no preference defined between $t_1$ and $t_3$ or $t_2$ and $t_3$,
the score of $t_3$ should be equal to the scores of both $t_1$ and $t_2$.
But this implies that the scores of $t_1$ and $t_2$ are equal
which is not possible since $t_1$ is preferred over $t_2$.
\end{example}
This lack of expressiveness of the quantitative approach is well known
in utility theory \cite{Fish70}. The paper \cite{ChTODS03} contains an extensive
discussion of the preference query literature.

In the earlier work on preference queries \cite{ChTODS03,Kie02}, one can find
positive and negative results about closure of different classes of orders,
including SPOs and WOs, under various composition operators. The results
in the present paper are, however, new. Restricting the relations
$\succ$ and $\succ_0$ (for example, assuming the interval order property
and compatibility)
and applying transitive closure where necessary make it possible to 
come up with positive counterparts of the negative results in \cite{ChTODS03}.
For example, \cite{ChTODS03} shows that SPOs and WOs are in general not closed w.r.t. union,
which should be contrasted with Theorems \ref{th:spo:union} and \ref{th:weak:weak}.
In \cite{Kie02}, Pareto and prioritized composition are defined somewhat differently
from the present paper. The operators combine  two preference relations, each defined over
some database relation. The resulting preference relation is defined over the 
Cartesian product of the database relations. So such operators are not useful in the
context of revision of preference relations.
On the other hand, the careful design of the language guarantees that every preference
relation that can be defined is an SPO.

Probably the most thoroughly studied class of qualitative preference queries
is the class of {\em skyline} queries.
A skyline query partitions all the attributes of a relation into DIFF, MAX, and
MIN attributes. Only tuples with identical values of all DIFF attributes are comparable;
among those, MAX attribute values are maximized and MIN values are minimized.
The query in Example \ref{ex:car}  is a very simple skyline query \cite{BoKoSt01},
with {\em Make} as a DIFF and {\em Year} as a MAX attribute.
Without DIFF attributes, a skyline is a special case of $n$-ary Pareto composition.

Various algorithms for evaluating qualitative preference queries are described
in \cite{ChTODS03,ToCi02}, and for evaluating skyline queries,
in \cite{BoKoSt01,PaTaFuSe03,BalEDBT04}.
\cite{BalVLDB04} describes how to implement preference queries that use
Pareto compositions of utility-based preference relations.
In Preference SQL \cite{KiKo02} general preference queries are implemented by a translation to SQL.
\cite{HrPa04} describes how materialized results of utility-based preference queries
can be used to answer other queries
of the same kind.

\subsection{CP-nets}

CP-nets \cite{BouetalJAIR04} are an influential recent formalism for reasoning with conditional
preference statements under {\em ceteris paribus} semantics (such semantics is also adopted in other work
\cite{McGDo04,WeDo91}).
We conjecture that CP-nets can be expressed in the framework of preference relations of \cite{ChTODS03}, used
in the present paper, by making the semantics explicit.
If the conjecture is true, the results of the present paper will be relevant to revision
of CP-nets.
\begin{example}
The CP-net \mbox{$M=\{a\succ \bar{a}, a:b\succ \bar{b}, \bar{a}: \bar{b}\succ b\}$}
where $a$ and $b$ are Boo\-lean variables,
captures the following preferences:
(1) prefer $a$ to $\bar{a}$, all else being equal;
(2) if $a$, prefer $b$ to $\bar{b}$; 
(3) if $\bar{a}$, prefer $\bar{b}$ to $b$.
We construct a preference relation $\succ_{C_M}$ between worlds, i.e., Boolean valuations of $a$ and $b$:
\[
\begin{array}{lcl}
(a,b)\succ_{C_M} (a',b') &\equiv &a=1 \wedge a'=0 \wedge b=b'\\&\vee & a=1 \wedge 
a'=1 \wedge b=1 \wedge b'=0 \\&\vee & a=0\wedge a'=0 \wedge b=0 \wedge b'=1.
\end{array}\]
Finally, the semantics of the CP-net is fully captured as the transitive closure $\;\tc{\succ_{C_M}}$.
Such closure can be computed using Constraint Datalog with Boolean constraints \cite{CDB00}.
\end{example}
CP-nets and related formalisms cannot express preference relations over infinite domains
which are essential in database applications.

\section {Conclusions and future work}\label{sec:concl}

We have presented a formal foundation for an iterative and incremental
approach to constructing ans evaluating preference queries. 
Our main focus is on {\em query modification}, a query transformation approach which works by revising the 
preference relation in the query. We have provided a detailed analysis of the cases
where the order-theoretic properties of the preference relation are preserved by the
revision. We have considered a number of different revision operators: union,
prioritized and Pareto composition.
We have also formulated algebraic laws that enable incremental evaluation of preference
queries.
Finally, we have studied the strengthening of the properties of preference relations
through finite restriction and weak-order extension.

Tables \ref{tab:refinement} and \ref{tab:overriding} summarize the closure properties of preference revision
under union and prioritized composition. There is no separate table for Pareto composition, because
there are only few results specific to this kind of composition.
\begin{table}[htb]
\centering
\begin{tabular}{|l|l|l|l|}
\hline
& $\succ$ SPO  &$\succ IO$ & $\succ$ WO \\\hline
$\succ_0$ SPO & not closed     & TC SPO if $0$-compat.  & SPO if $0$-compat.\\\hline
$\succ_0$ IO & TC SPO if $0$-compat. & TC SPO if $0$-compat.&SPO if $0$-compat. \\\hline
$\succ_0$ WO &SPO if $0$-compat. & SPO if $0$-compat.& WO if $0$-compat.\\\hline
\end{tabular}
\caption{Revision using union}
\label{tab:refinement}
\end{table}
\begin{table}[htb]
\centering
\begin{tabular}{|l|l|l|l|}
\hline
&$\succ$ SPO &$\succ IO$ & $\succ$ WO \\\hline
$\succ_0$ SPO&not closed &TC SPO if $0$-compat. & SPO if $0$-compat.\\\hline
$\succ_0$ IO & TC SPO if $1$-compat. & TC SPO if $1$-compat. & TC SPO if $1$-compat.\\\hline
$\succ_0$ WO & SPO & SPO &WO\\\hline
\end{tabular}
\caption{Revision using prioritized composition}
\label{tab:overriding}
\end{table}

Future work includes the integration of our results with standard query optimization techniques,
both rewriting- and cost-based.
Semantic query optimization techniques for preference queries \cite{ChCDB04} can also
be applied in this context.
Another possible direction could lead to the design of a {\em revision language}
in which richer classes of preference revisions can be specified \cite{GrMeRe97}.

One should also consider possible courses of action if the original preference relation
$\succ$ and $\succ_0$ lack the property of compatibility, for example if $\succ$ and $\succ_0$
are not $0$-compatible in the case of revision by union.
Then the target of the revision is an SPO which is the closest to the preference relation $\succ\cup\succ_0$.
Such an SPO will not  be unique. Moreover, it is not clear how to obtain ipfs defining the
revisions. Similarly, one could study {\em contraction} of preference relations. The need
for contraction arises, for example,  when a user realizes that the result of a preference query
does not contain some expected tuples.

Finally, one can consider preference query transformations which go beyond preference
revision, as well as more general classes of preference queries that involve, for example,
ranking \cite{ChTODS03}.

\bibliography{../main}
\newcommand{\etalchar}[1]{$^{#1}$}

\end{document}